\documentclass[letterpaper,twocolumn,10pt]{article}
\usepackage{usenix_style}

\makeatletter
\def\@maketitle{\newpage
  \vbox to 1.0in{%
    \vspace*{\fill}%
    \begin{center}%
      {\Large\bfseries \@title \par}%
      \vskip 0.1in minus 0.05in
      {\large\itshape
      \lineskip .3em
      \begin{tabular}[t]{c}\@author\end{tabular}\par}%
    \end{center}%
    \vspace*{\fill}%
  }%
}
\makeatother

\usepackage{tikz}
\usepackage{amsmath}

\usepackage{filecontents}
\usepackage{soul}

\usepackage{algorithm}  
\usepackage{algorithmicx}  
\usepackage[noend]{algpseudocode}
\usepackage{eqparbox}
\usepackage{xcolor}

\usepackage{graphicx}
\usepackage{subfigure}
\usepackage{enumitem}

\usepackage[hang,flushmargin]{footmisc}

\usepackage[skip=0pt]{caption} 
\usepackage{tabularx}
\usepackage{booktabs}

\usepackage{soul}

\usepackage{url}
\usepackage{multirow}
\usepackage{comment}
\usepackage[explicit]{titlesec}

\usepackage{pifont}
\usepackage{adjustbox}
\usepackage{todonotes}
\usepackage{latexsym}

\begin{document}

\title{\textsc{Lotus}: Optimizing Disaggregated Transactions with Disaggregated Locks}
\author{
Zhisheng Hu$^{1}$ \enspace
Pengfei Zuo$^{2}$ \enspace
Junliang Hu$^{1}$ \enspace
Yizou Chen$^{1}$ \enspace
Yingjia Wang$^{1}$ \enspace
Ming-Chang Yang$^{1}$ \\
$^{1}$The Chinese University of Hong Kong \quad
$^{2}$Huawei
}

\maketitle

\begin{abstract}
Disaggregated memory (DM) separates compute and memory resources, allowing flexible scaling to achieve high resource utilization. To ensure atomic and consistent data access on DM, distributed transaction systems have been adapted, where compute nodes (CNs) rely on one-sided RDMA operations to access remote data in memory nodes (MNs). However, we observe that in existing transaction systems, the RDMA network interface cards at MNs become a primary performance bottleneck. This bottleneck arises from the high volume of one-sided atomic operations used for locks, which hinders the system's ability to scale efficiently.

To address this issue, this paper presents \textsc{Lotus}, a scalable distributed transaction system with lock disaggregation on DM. The key innovation of \textsc{Lotus} is to disaggregate locks from data and execute all locks on CNs, thus eliminating the bottleneck at MN RNICs. To achieve efficient lock management on CNs, \textsc{Lotus} employs an \textit{application-aware lock management} mechanism that leverages the locality of the OLTP workloads to shard locks while maintaining load balance. To ensure consistent transaction processing with lock disaggregation, \textsc{Lotus} introduces a \textit{lock-first transaction protocol}, which separates the locking phase as the first step in each read-write transaction execution. This protocol allows the system to determine the success of lock acquisitions early and proactively abort conflicting transactions, improving overall efficiency. To tolerate lock loss during CN failures, \textsc{Lotus} employs a \textit{lock-rebuild-free recovery} mechanism that treats locks as ephemeral and avoids their reconstruction, ensuring lightweight recovery for CN failures. Experimental results demonstrate that \textsc{Lotus} improves transaction throughput by up to 2.1$\times$ and reduces latency by up to 49.4\% compared to state-of-the-art transaction systems on DM.
\end{abstract}

\section{Introduction}\label{section_introduction}
\label{section_introduction_chall}
Disaggregated memory (DM) has recently gained significant attention from both
industry~\cite{polarDB, aurora, cxl-cloud-native, gaussdb} and academia~\cite{infiniswap, bladeser, legoos, semeru, cxl-shm}. It decouples compute and memory resources from traditional monolithic servers, forming independent resource pools. Typically, a DM architecture consists of a compute pool comprising multiple compute nodes (CNs) with powerful CPUs but limited local memory, and a memory pool comprising memory nodes (MNs) with substantial memory but weak CPUs, interconnected via high-speed networks such as RDMA~\cite{rdmaspec} or CXL~\cite{cxlspec}. This architecture allows for flexible allocation and efficient sharing of resources across multiple nodes, enhancing overall resource utilization. 

To provide atomic and consistent data access services for applications running on DM, distributed transaction systems have been adapted to DM. Examples include FORD~\cite{ford}, which employs a single-versioning design, and Motor~\cite{motor}, which enables multi-versioning concurrency control (MVCC). 
Both systems place transaction processing on CNs while storing data in the memory of MNs, with CNs leveraging one-sided RDMA to access data and bypass MN CPUs. Particularly, CNs rely on one-sided \texttt{RDMA} \texttt{Atomic} operations, such as \texttt{CAS} and \texttt{FAA}, to lock data during concurrency control.

However, these transaction systems on DM face a significant bottleneck at the RDMA network interface cards (RNICs) of MNs. First, \texttt{RDMA} \texttt{Atomic} operations (\textit{i.e.}, \texttt{CAS} and \texttt{FAA}) typically incur higher overhead than \texttt{RDMA} \texttt{READ} and \texttt{WRITE} due to their complex semantics~\cite{understanding}, limiting performance more easily. The maximum IOPS of RNICs for atomic operations is significantly lower than that of \texttt{READ} and \texttt{WRITE} operations~\cite{understanding, design-guide-high, hybrid-better, rdmaturing}. 
Moreover, when a transaction involves writing to multiple data records, it requires multiple \texttt{CAS} operations to lock them, followed by multiple \texttt{READ}s and \texttt{WRITE}s. As the number of concurrent transactions increases, the processing capacity of MN RNICs can be easily overwhelmed, becoming IOPS-bound.
More critically, in existing transaction systems, when lock acquisition fails for one data record, the entire transaction aborts, necessitating the release of all previously acquired locks. This process results in substantial waste of network bandwidth and processing resources.

To address this problem, this paper proposes that lock disaggregation is an effective way to mitigate the network bottleneck of MNs in existing systems. Specifically, we decouple locks from their data: locks are placed in the compute pool, while data is in the memory pool. This allows lock operations to be handled entirely within the compute pool, avoiding access to the memory pool and mitigating the bottleneck of MN RNICs.
Locks are distributed across CNs, with each CN responsible for a specific range. All lock operations for keys within that range are handled by the corresponding CN, where the CPUs process the locks, eliminating the excessive use of \texttt{RDMA} \texttt{Atomic} operations to MNs. When the locks required by a transaction are managed locally on the corresponding CN, many lock operations can be performed locally, further improving system performance.

However, fully leveraging lock disaggregation presents several challenges.
\textit{1) Scattered lock management on CNs.} Distributing locks across CNs introduces inefficiencies, as transactions may require multiple requests to different CNs to lock and unlock keys, resulting in high communication overhead and increased latency~\cite{tail-scale}. Additionally, workload skewness can cause lock load imbalance, where certain CNs handle disproportionately more locking operations, further degrading system performance.
\textit{2) Complex consistency guarantee.} In traditional transaction systems, coordination occurs only between CNs and MNs. However, with lock disaggregation, additional coordination across multiple CNs is required to maintain consistency, making the consistency guarantee more complex. \textit{3) Lock loss during CN failures.} Unlike locks stored in MNs, which can benefit from replication for higher reliability, locks in CNs are more prone to failure. Therefore, ensuring fault tolerance in CNs becomes crucial for maintaining the overall robustness and reliability.

To address the above challenges, we propose \textbf{\textsc{Lotus}}, a scalable distributed transaction system with lock disaggregation on disaggregated memory. \textsc{Lotus} incorporates the following key techniques.

First, to achieve efficient lock management on CNs, \textsc{Lotus} employs an \textit{application-aware lock management} mechanism that leverages the locality of the OLTP workloads to shard locks with load balance. Specifically, this mechanism shards the locks required by each transaction on a few or even a single CN, allowing for efficient batched locking to mitigate the high communication overhead across CNs. By further applying a two-level balancing strategy, this mechanism enables most lock operations to be performed locally while maintaining load balance among CNs. 

Second, to ensure consistent transaction processing, \textsc{Lotus} designs a \textit{lock-first transaction protocol} built on top of lock disaggregation. 
A key feature of this protocol is the separation of the locking phase as the first step in each read-write transaction execution. By determining the success of lock acquisitions early, it proactively aborts conflicting transactions, thus preventing unnecessary performance degradation.

Third, to tolerate lock loss during CN failures, \textsc{Lotus} employs a \textit{lock-rebuild-free recovery} mechanism. In the event of a CN failure, transaction recovery proceeds independently of CN recovery, with \textsc{Lotus} aborting transactions waiting for locks from the failed CN and allowing transactions already in the commit phase to complete safely.
By treating locks as ephemeral and avoiding their reconstruction, \textsc{Lotus} ensures correct and lightweight recovery for CN failures.

We implement \textsc{Lotus} and evaluate its effectiveness using \texttt{KVS}, \texttt{TATP}, \texttt{SmallBank}, and \texttt{TPCC} benchmarks. The results show that, compared with the state-of-the-art transaction systems on DM, \textsc{Lotus} improves transaction throughput by up to 2.1$\times$ and reduces latency by up to 49.4\%. 

In summary, this paper makes the following contributions:
\begin{itemize}[leftmargin=20pt, itemsep=0pt, topsep=2pt]
  \item We identify a critical issue in existing transaction systems on DM, \textit{i.e.}, network congestion at MN RNICs caused by the excessive use of one-sided \texttt{RDMA} \texttt{Atomic} operations. We propose that this issue can be mitigated through lock disaggregation.
  \item We propose \textsc{Lotus}, a distributed transaction system with lock disaggregation on DM. By decoupling locks from data, \textsc{Lotus} processes locks entirely within the compute pool, bypassing the memory pool and eliminating the bottleneck at MN RNICs. Furthermore, \textsc{Lotus} incorporates a series of design innovations to address key challenges and fully benefit from lock disaggregation.
  \item We implement \textsc{Lotus} and conduct extensive evaluations, demonstrating the effectiveness of our design.
\end{itemize}
\section{Background and Motivation}

\begin{figure}[tp]
    \centering
    \includegraphics[width=\linewidth]{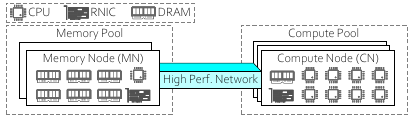}
    \captionsetup{font=small}
    \caption{The architecture of disaggregated memory.}
    \label{FIG-BGROUND}
\end{figure}

\subsection{Disaggregated Memory}\label{section_bg_dm}

Disaggregated memory (DM) has recently garnered significant attention from both industry and academia~\cite{polarDB, aurora, cxl-cloud-native, bladeser, legoos, semeru, cxl-shm, zombie}. Its design philosophy revolves around the decoupling of compute and memory, where compute nodes (CNs) with strong CPUs but limited memory are placed in a compute pool, while memory nodes (MNs) with abundant memory but weak CPUs are placed in a memory pool, as shown in Figure~\ref{FIG-BGROUND}. Resources are dynamically allocated based on application requirements, minimizing resource waste. The concept of DM is also being applied in other domains, such as LLM inferencing, including the disaggregation of compute-intensive and memory-intensive tasks to improve GPU utilization~\cite{distserve, splitwise, mooncacke} and the sharing of KV cache in the memory pool to avoid redundant computation~\cite{memserve, cachedatten}.

In DM architectures, nodes are interconnected via high-performance networks, including established technologies like RDMA~\cite{rdmaspec}, as well as emerging ones such as CXL~\cite{cxlspec, cxlpond, cxl-cloud-native, oasis} and NVLink~\cite{nvlinkspec}. This paper focuses on RDMA, which is widely adopted in DM researches~\cite{rolex, aceso, fusee, rethink, polarDB}. RDMA offers one-sided and two-sided verbs for communication. One-sided verbs enable threads in CNs to directly read and write memory in MNs without involving the MN CPUs, whereas two-sided verbs require the participation of  CPUs on both sides, similar to traditional TCP/IP socket communication. Notably, one-sided verbs include atomic operations like \texttt{CAS} and \texttt{FAA}, whose atomicity is extensively utilized on DM to ensure concurrency control~\cite{design-guide-sync}.

\subsection{Distributed Transactions on DM}\label{section_back_ft}

\textbf{System Model.} To provide atomic and consistent multi-data read/write services for applications on DM~\cite{ford, motor}, distributed transaction systems are ported to DM. Coordinators, which execute transactions, run on CNs, while data is stored in MNs and accessed directly by coordinators via one-sided RDMA. A small amount of memory in CNs can be utilized as a cache to accelerate transaction execution, such as caching remote data addresses. 
To ensure fault tolerance for data in MNs, replication techniques like primary-backup replication~\cite{primary-backup} are commonly employed. 

\begin{figure}[tbp]
    \centering
    \begin{minipage}[t]{0.49\linewidth}
      \centering
      \includegraphics[width=\textwidth]{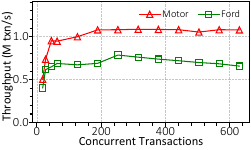}
      \captionsetup{font=small}
      \caption{The throughput of Motor and FORD on \texttt{SmallBank}.}
      \label{FIG-EXP-PRE-TPT}
    \end{minipage}
    \hspace{0em}
    \begin{minipage}[t]{0.49\linewidth}
      \centering
      \includegraphics[width=\textwidth]{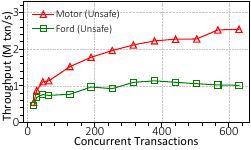}
      \captionsetup{font=small}
      \caption{The throughput of Motor and FORD on \texttt{SmallBank} after abandoning \texttt{CAS}.}
      \label{FIG-EXP-PRE-TPT-NOCAS}
    \end{minipage}
\end{figure}

\noindent \textbf{Existing Approaches.} Currently, the state-of-the-art transaction systems on DM include FORD~\cite{ford} and Motor~\cite{motor}. FORD is a single-versioned transaction system whose protocol is entirely executed using one-sided RDMA operations. It employs optimizations such as doorbell-batched \texttt{RDMA} \texttt{CAS}+\texttt{READ} to lock and read data within a single round-trip time (RTT). Motor, an evolution of FORD, introduces multi-version concurrency control (MVCC), thereby achieving higher concurrency. 

\noindent \textbf{Limitations of Existing Approaches.}
Existing DM transaction systems suffer from a critical flaw: a large number of one-sided atomic operations can easily overwhelm the processing capacity of MN RNICs. A single transaction may trigger numerous RDMA requests, such as fetching addresses from indexes, acquiring locks, and reading or writing records. 
Particularly, \texttt{RDMA} \texttt{Atomic} operations used for locking (\textit{e.g.}, \texttt{CAS}) generally incur higher overhead than \texttt{RDMA} \texttt{READ}/\texttt{WRITE} operations due to their more complex implementations, a characteristic observed across various RNIC types (\textit{e.g.}, ConnectX-3/4/5/6)~\cite{understanding, design-guide-high, hybrid-better, rdmaturing}.
For example, using Perftest~\cite{perftestspec}, we measure that in our experimental setup, a single RNIC acting as the remote side can handle up to 35 Mops for \texttt{RDMA} \texttt{WRITE} (8B) requests, while the maximum IOPS for \texttt{CAS} (8B) is only 2.5 Mops. In typical DM transaction systems, where a single MN may serve multiple coordinators in an N-to-1 configuration, this can overwhelm the MN RNICs' processing capacity easily. A more severe issue is that when a transaction fails to acquire a lock, it aborts, wasting significant network resources already expended on successfully locked data.

To illustrate this issue, we evaluate Motor and FORD using the \texttt{SmallBank} benchmark under the same setup in \S\ref{section_exp_setup}. As shown in Figure~\ref{FIG-EXP-PRE-TPT}, by increasing the number of concurrent transactions, we observe that 3 MNs can only efficiently serve up to 45 concurrent transactions before hitting a bottleneck, beyond which latency increases significantly. 
Overall, current state-of-the-art transaction systems on DM remain inefficient, leaving substantial room for improvement. 

\subsection{Observation and Motivation}\label{section_background_locality}

\begin{figure}[tp]
    \centering
    \includegraphics[width=\linewidth]{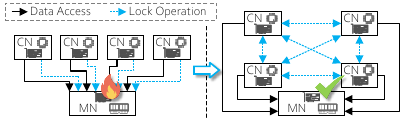}
    \captionsetup{font=small}
    \caption{Decoupling data and locks on disaggregated memory.}
    \label{FIG-MOTIVATION}
\end{figure}

\noindent \textbf{Observation.}
In Figure~\ref{FIG-EXP-PRE-TPT-NOCAS}, we abandon the use of \texttt{CAS} in Motor and FORD (\textit{i.e.}, operating unsafely). The transaction abort rates remain largely unchanged, as we use a sufficiently large dataset (\textit{i.e.}, 20M accounts). 
The results show that both Motor and FORD achieve significant performance gains. Notably, Motor’s throughput continues to scale with increasing concurrency, reaching 2.6 million transactions per second, a 2.4$\times$ improvement over the previous peak (Figure~\ref{FIG-EXP-PRE-TPT}). FORD shows smaller gains because its hash buckets store versions together with full values, causing it to become bandwidth-bound early. Nevertheless, the overall performance surge in both systems highlights the benefit of removing lock operations from the network path between CNs and MNs.

\noindent \textbf{Motivation.} 
Based on this observation, we argue that disaggregating locks to CNs can address the limitations of existing approaches. Specifically, although CNs have limited memory, it is sufficient to store small-sized locks. Locks can be distributed across CNs, with each CN responsible for managing a specific range of keys. All lock operations for keys within that range are handled by the corresponding CN, effectively bypassing the memory pool, as shown in Figure~\ref{FIG-MOTIVATION}. 
In this way, MN RNICs are relieved of the heavy burden of \texttt{RDMA} \texttt{Atomic} operations, freeing up significant network resources to focus on handling data transfer requests, improving performance.

\noindent \textbf{Challenges.}\label{section_chall}
However, as analyzed in \S\ref{section_introduction_chall}, applying lock disaggregation to DM introduces several challenges. 1) Distributing locks to CNs increases communication overhead and latency, as transactions may need to contact multiple CNs to lock or unlock keys. Workload skewness further exacerbates this by creating performance hotspots on overloaded CNs. 2) Maintaining consistency becomes more complex, as additional coordination across CNs is required. 3) Storing locks in CNs makes fault tolerance critical, since losing locks due to CN failures can compromise the system’s reliability.

\section{The System Overview of \textsc{Lotus}}\label{section_design_overview}

To address the aforementioned challenges, 
we propose \textsc{Lotus}, which enables efficient lock disaggregation for memory-disaggregated transactions.

\begin{figure}[tp]
    \centering
    \includegraphics[width=\linewidth]{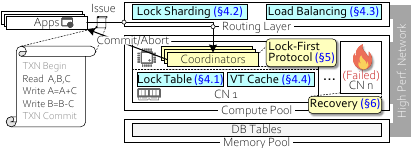}
    \captionsetup{font=small}
    \caption{The overview of \textsc{Lotus}.}
    \label{FIG-OVERVIEW}
\end{figure}

\noindent \textbf{Overview.} Figure~\ref{FIG-OVERVIEW} illustrates the overview of \textsc{Lotus}. In the compute nodes (CNs), \textit{coordinators} are threads that execute transactions. The \textit{lock table} stores the locks managed by the CN (\S\ref{section_design_shardlock}), while the \textit{version table cache} (\textit{i.e.}, \textit{VT cache}) stores version tables of data to expedite data retrieval (\S\ref{section_design_vtcache}). Memory nodes (MNs) house the \textit{DB tables}, which are directly accessed by the coordinators via one-sided RDMA. 
Transactions are routed to a CN via a routing layer that employs \textit{application-aware lock sharding} (\S\ref{section_design_sharding}) and \textit{two-level load balancing} (\S\ref{section_design_balancing}).
Transactions are executed following a \textit{lock-first transaction protocol} (\S\ref{section_protocol}). CN failures are handled with the \textit{lock-rebuild-free recovery} mechanism (\S\ref{section_cn_recovery}).

\noindent \textbf{Workflow.} \textit{1) Init phase.} Upon \textsc{Lotus} startup, MNs utilize their limited CPUs to allocate memory and register it as RDMA regions. Subsequently, application data is loaded into DB tables, which are organized by indexes (see \S\ref{section-mem-store}). CNs then establish RDMA reliable connected (RC) QP connections with MNs, and MNs send DB table metadata to CNs for caching. This metadata (only a few KB), \textit{e.g.}, the start addresses of DB tables, helps coordinators correctly locate data. CNs communicate with each other through unreliable datagram (UD) QP based RPCs~\cite{FaSST}, with a timeout mechanism ensuring reliability. 
\textit{2) Run phase.} During \textsc{Lotus} operation, upper-layer applications issue transactions to CNs via a routing layer. Coordinators execute these transactions according to the transaction protocol. Generally, the process involves locking the data, reading it, and processing it based on the transaction logic. Coordinators then commit updates to the memory pool and unlock the data. Throughout this process, data locking and unlocking are performed entirely within the compute pool, bypassing the memory pool. 

\section{Application-Aware Lock Management}

\textsc{Lotus} proposes lock disaggregation to address the limitations of existing approaches.
However, managing locks across CNs is challenging. Coordinators may frequently communicate with a large number of CNs to lock data, and workload imbalance can overwhelm certain CNs with disproportionate lock requests.
To address these, \textsc{Lotus} employs an \textit{application-aware lock management}. First, \textsc{Lotus} maintains a lock table in each CN to efficiently manage locks (\S\ref{section_design_shardlock}). Second, \textsc{Lotus} leverages application-specific knowledge to shard locks, supporting batched locking (\S\ref{section_design_sharding}). Third, \textsc{Lotus} adopts a two-level strategy to distribute transactions to CNs, facilitating local locking while maintaining load balance (\S\ref{section_design_balancing}). Finally, building on the transaction routing strategy, \textsc{Lotus} keeps a version table cache in each CN to accelerate data retrieval with zero consistency overhead (\S\ref{section_design_vtcache}).

\subsection{Distributed Lock Table}\label{section_design_shardlock}

\begin{figure}[tp]
    \centering
    \includegraphics[width=\linewidth]{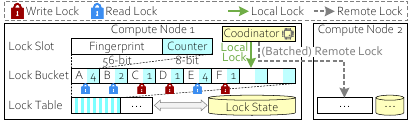}
    \captionsetup{font=small}
    \caption{The distributed lock table in \textsc{Lotus}.}
    \label{FIG-LOCKS}
    \vspace{-0.4em}
\end{figure}

\begin{algorithm}[tbp]
\footnotesize
\caption{\textsc{Lotus} lock algorithm.}\label{alg_lock}
\begin{algorithmic}[1]
\State \textcolor[RGB]{128, 128, 128}{$is\_write \gets \text{write lock flag (1 for write, 0 for read)}$} 
\State \textcolor[RGB]{128, 128, 128}{$cn\_id \gets \text{compute node ID (which CN the transaction is running on)}$}
\State \textcolor[RGB]{128, 128, 128}{$txn\_id \gets \text{transaction ID}$}

\Procedure{\textcolor{blue}{Acquire}}{$key,\ is\_write,\ cn\_id,\ txn\_id$} $\to\ \textbf{bool}$
    \If{$\textcolor{blue}{\textsc{CheckLockState}}(key,\ is\_write,\ cn\_id, txn\_id)$}\label{line_check_lock_state}
        \State \Return \texttt{True} \textcolor[RGB]{128, 128, 128}{\Comment{\textit{This key is already locked by the transaction}}}
    \EndIf
    \State $fingerprint,\ bucket \gets \textcolor{blue}{\textsc{Hash}}(key)$
    \State $slot \gets bucket.\textcolor{blue}{\textsc{FindMatch}}(fingerprint)$\label{line_find_match}
    \If{$slot = \texttt{Null}\ \textbf{or}\ slot.\textcolor{blue}{\textsc{IsConflictLocked}}(is\_write)$}
        \State \Return \texttt{False} \textcolor[RGB]{128, 128, 128}{\Comment{\textit{Return if any lock condition is not met}}}\label{line_conflict_locked}
    \EndIf
    \If{$is\_write$} \textcolor[RGB]{128, 128, 128}{\Comment{\textit{Acquire write lock}}}
        \State $compare \gets 0$
        \State $swap \gets \{ fingerprint,\ 1 \}$
        \If{$\textcolor{blue}{\textsc{NotLocal}}(cn\_id)$} 
            \State $\textcolor{blue}{\textsc{InvalidateCache}}(key)$ \textcolor[RGB]{128, 128, 128}{\Comment{\textit{Invalidate VT cache (\S4.4)}}}\label{line_invalid_cache}
        \EndIf
    \Else \textcolor[RGB]{128, 128, 128}{\Comment{\textit{Acquire read lock}}}
        \State $compare \gets slot.\textcolor{blue}{\textsc{Get}}()$
        \State $swap \gets \{ fingerprint,\ slot.\texttt{Counter} + 2 \}$
    \EndIf
    
    \State $ret \gets slot.\textcolor{blue}{\textsc{CompareAndSwap}}(compare,\ swap)$
    \If{$ret$} \textcolor[RGB]{128, 128, 128}{\Comment{\textit{On success, log this operation to lock state}}}
        \State $\textcolor{blue}{\textsc{UpdateLockState}}(key,\ is\_write,\ cn\_id, txn\_id)$\label{line_update_lock_state}
    \EndIf
    \State \Return $ret$
\EndProcedure
\end{algorithmic}
\end{algorithm}

\textsc{Lotus} achieves lock disaggregation by distributing locks to CNs, with each CN maintaining its own lock table.
As shown in Figure~\ref{FIG-LOCKS}, the lock table in each CN is implemented as a fixed-length hash table, which supports both read and write locks to enable fine-grained concurrency control (see \S\ref{section_protocol}). Each slot in the hash table is 8B in size, consisting of a 7B \textit{fingerprint} (generated from the hash of the key) and a 1B \textit{counter}. Every 8 slots form a \textit{lock bucket}. 
The lock table also maintains a mapping in the \textit{lock state} that records, for each held lock, its holders' transaction IDs, CN IDs, and lock modes, which is used to reject redundant lock requests, ensuring lock operation idempotency.

To lock a data record, the lock request is sent to a CN based on the shard-to-CN mapping (see \S\ref{section_design_sharding}), where the lock operation is executed (Algorithm~\ref{alg_lock}). The CN first checks if the transaction already holds the lock (line~\ref{line_check_lock_state}), then hashes the key to obtain its fingerprint and lock bucket. The fingerprint matches a lock slot (line~\ref{line_find_match}), whose counter encodes the lock state: 1 for a write lock, nonzero even for read locks. 
If the bucket is full (\textit{i.e.}, slot is \texttt{null}), or the slot is locked by a conflicting lock (\textit{i.e.}, read-write or write-write conflicts), or acquiring a read lock would cause the counter to overflow, the attempt fails (line~\ref{line_conflict_locked}). 
For a write lock request from a remote node, the CN also invalidates the corresponding entry in the version table cache (line~\ref{line_invalid_cache}, see \S\ref{section_design_vtcache}).
Finally, a \texttt{CAS} instruction is used to atomically acquire the lock. Upon success, the lock state is updated (line~\ref{line_update_lock_state}). 
Unlocking follows a similar procedure, decrementing the counter as appropriate. 

Notably, for insert operations, locking is required for both the key and the bucket address in the remote index (referred to as the \textit{index bucket}).
This is achieved by using the index bucket address as a key to locate the lock in lock tables. By locking the index bucket, \textsc{Lotus} ensures concurrency safety for insert operations. If another transaction attempts to insert into the same index bucket, it will fail to acquire the lock. 

When coordinators process transactions, they handle lock requests as follows: if the lock is local (\textit{i.e.}, within the managing scope of the local CN), it is directly executed in the local lock table. For remote lock requests (\textit{i.e.}, within the scope of another CN), the request is forwarded to the corresponding CN via \texttt{RDMA} \texttt{SEND}/\texttt{RECV} based RPC, where the remote coordinator executes the lock operation and replies with the result. To optimize efficiency, multiple remote lock requests targeting the same CN from a transaction are batched into a single RDMA message, saving IOPS. 
In our current implementation, each CN has an equal number of coordinators (\textit{i.e.}, threads), and the i-th coordinator in one CN only sends RPCs to the i-th coordinator in another CN, achieving simple load balancing and avoiding additional CPU consumption.

\subsection{Application-Aware Lock Sharding}\label{section_design_sharding}

\begin{figure}[tp]
    \centering
    \includegraphics[width=\linewidth]{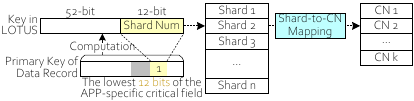}
    \captionsetup{font=small}
    \caption{The application-aware lock sharding in \textsc{Lotus}.}
    \label{FIG-PARTITION}
\end{figure}

After establishing a lock table in each CN, it is crucial to partition data locks across CNs, so that each CN is responsible for a specific subset of locks. However, this poses a challenge: if the locks are poorly partitioned, a read-write transaction can trigger lock requests to a large number of CNs, which degrades performance~\cite{tail-scale}. 
To address this, \textsc{Lotus} employs application-aware lock sharding, leveraging OLTP workload locality to ensure the locks required by each read-write transaction are concentrated on a few or even a single CN. This enables batched locking and reduces coordination overhead. To achieve this, \textsc{Lotus} first divides data into multiple shards and then establishes a mapping between shards and CNs.

\noindent \textbf{OLTP Workload Locality.}
In OLTP workloads, transaction access patterns often exhibit strong locality, a phenomenon widely observed in real-world applications.  
For instance, over 85\% of \texttt{TPCC} transactions~\cite{tpccspec} occur within a single warehouse; 85\% of user data accesses in the Yahoo! trace~\cite{yahootrace} exhibit regional locality; and payment system users~\cite{venmo} transact predominantly within a small set of friend accounts, demonstrating pronounced locality. 
While legacy distributed transaction systems~\cite{zeus, non-2pc, tigatxn} have leveraged locality to enhance performance, exploiting it on DM remains unexplored. 

\noindent \textbf{Partitioning Data into Shards.}
\textsc{Lotus} uses an application-specific hash function to generate a 64-bit key for indexing each data record based on its primary key, consistent with previous systems~\cite{fusee, motor, ford}. The primary key, which can be a single field or a combination of fields, uniquely identifies each record in a DB table.
The hash function is specified by the user (or application developer) for each DB table prior to deployment.
In \textsc{Lotus}, when specifying the hash function for a DB table, the user can also specify a field with the highest locality (\textit{i.e.}, the critical field) based on application semantics (if not specified, data is sharded randomly). This field is then used to shard the data, ensuring that most locality-sensitive records remain within the same shard. 
As illustrated in Figure~\ref{FIG-PARTITION}, the function uses the lowest 12 bits of the critical field in the primary key as the lowest 12 bits of the \textsc{Lotus} key, referred to as the \textit{shard number}. The remaining 52 bits are derived from all fields of the primary key through a predefined computation, ensuring the uniqueness of each key within the DB table. \textsc{Lotus} uses the shard number as the basis for partitioning data into shards.
Each CN is responsible for one or more shards. In this way, many locks for data records with locality can be placed in the same shard. For example, in \texttt{TPCC}, the lowest 12 bits of the warehouse ID serves as the shard number, ensuring that lock requests for data within the same warehouse are handled by a single CN. Similarly, in \texttt{TATP}, the subscriber ID ensures that most transactions involving a single subscriber are processed within one CN. This provides opportunities for efficient batched locking.

Note that this design aligns with standard deployment practices in modern OLTP systems, such as MongoDB's shard keys~\cite{mongodbspec}, and DynamoDB/CosmosDB/Cassandra's partition keys~\cite{dynamodbspec, cosmosdbspec, cassandraspec}, where users specify keys for locality and scalability. By simply reusing these existing keys as the critical field, \textsc{Lotus} leverages application semantics without introducing new tuning knobs. We evaluate the sensitivity to critical field choice, including suboptimal selections, in \S\ref{section_exp_sensitivity}.

\noindent \textbf{Mapping Shards to Compute Nodes.}
Transactions issued from upper-layer applications are routed to the appropriate CNs via a routing layer that caches the latest shard-to-CN mapping. 
We assume the routing layer is scalable and incurs no additional overhead to \textsc{Lotus}, as it's a common and essential infrastructure in industrial distributed databases~\cite{polarDB, aurora, dynamo, spanner} for request distributing. The specific design of the routing layer is orthogonal to our work.
Initially, the key range for lock management is evenly distributed among CNs. When extreme skewness arises, adaptive load balancing is performed (see \S\ref{section_design_balancing}). Each CN is aware of the key range of locks it manages. If it receives a lock request outside its managed range, it returns an error, prompting the requester to retry based on the latest shard-to-CN mapping.

\subsection{Two-Level Load Balancing}\label{section_design_balancing}

\begin{figure}[tp]
    \centering
    \includegraphics[width=\linewidth]{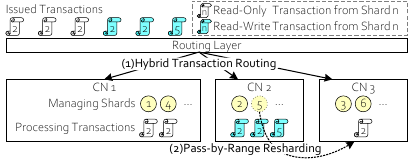}
    \captionsetup{font=small}
    \caption{The two-level load balancing in \textsc{Lotus}.}
    \label{FIG-ROUTING}
\end{figure}

\textsc{Lotus} adopts a two-level load balancing strategy, \textit{i.e.}, hybrid transaction routing and pass-by-range resharding
, to determine the rules for routing transactions to CNs. 

\noindent \textbf{Hybrid Transaction Routing.}
In \S\ref{section_design_sharding}, \textsc{Lotus} has assigned locks for data records with locality to the same CN by leveraging shard numbers. This ensures a high concentration of locks from a read-write transaction, \textit{i.e.}, the locks are centralized on a few or even a single CN, facilitating batched locking.
However, without a suitable transaction routing design, frequent remote lock requests can still degrade performance. To address this, \textsc{Lotus} employs a hybrid routing strategy that routes each read-write and read-only transaction to an appropriate CN, ensuring most lock requests are handled locally while mitigating load imbalance.

As shown in Figure~\ref{FIG-ROUTING}, \textsc{Lotus}'s hybrid routing strategy directs read-only transactions to a CN selected in a uniformly random manner, while read-write transactions are routed to the CN responsible for the shard corresponding to the shard number of the first data record in the transaction. 
This strategy aligns naturally with DM. Read-only transactions, which do not require locks, can efficiently access data in the memory pool on any CN, with random routing balancing the load. In contrast, read-write transactions, which involve significant writes, are routed to the CN responsible for their shards, ensuring most lock requests are handled locally to improve performance. While multiple read-only transactions on the same key can execute in parallel across nodes, read-write transactions on the same key cannot. Routing such transactions to a specific node enables early abortion of unnecessary ones, preventing performance waste.

\noindent \textbf{Pass-by-Range Resharding.}
Although the hybrid transaction routing strategy ensures a certain degree of load balancing, the inherent skewness in OLTP workloads can still cause a CN to become a performance bottleneck if it manages overly "hot" shards. To address this, \textsc{Lotus} employs pass-by-range resharding, which disperses hot shards across multiple CNs to achieve load balancing. 
Specifically, each CN periodically measures its transaction execution latency and calculates metrics such as the average latency, writing these metrics to a pre-allocated region in the memory pool every fixed interval (\textit{e.g.}, 100ms in our implementation). Each CN also periodically reads the performance metrics of other CNs for comparison. 
A CN is considered overloaded if its latency remains more than 50\% higher than the cluster-wide average for three consecutive intervals (300ms in total). Once overloaded, the CN identifies its hottest shard based on recent request rates and transfers its ownership to the CN with the lowest latency. 
This resharding is highly efficient on DM as it does not require transferring the data but only the ownership of locks.

During shard management changes, two CNs (the shard sender and the receiver) coordinate to ensure atomicity thro-ugh a straightforward mechanism. The sender first stops serving lock requests for the shard and waits for all previous locks to be unlocked (if waiting too long, \textit{e.g.}, 10ms, it proactively aborts the running transactions using the transaction and CN IDs recorded in the lock state), 
deletes stale metadata related to the shard, and clears the corresponding data in the local cache (\S\ref{section_design_vtcache}), as this shard will be managed by the receiver. It then sends a message to the receiver via \texttt{RDMA} \texttt{SEND}/\texttt{RECV}. Upon receiving the message, the receiver updates its managed shard range and begins serving lock requests for this shard. Finally, the sender updates the routing layer with the new shard mapping.
During the above procedure, the lock service for this shard will be briefly interrupted. However, since resharding occurs rarely (as workload access patterns change relatively slowly~\cite{scale-store, twitter-workload, pegasus, facebook-workload}), and the interruption is short (0.19ms for \texttt{TATP}, 0.86ms for \texttt{TPCC}, and 4.67ms for \texttt{SmallBank} in our experiments), we consider this acceptable.

\subsection{Version Table Cache}\label{section_design_vtcache}

\begin{figure}[tp]
    \centering
    \includegraphics[width=\linewidth]{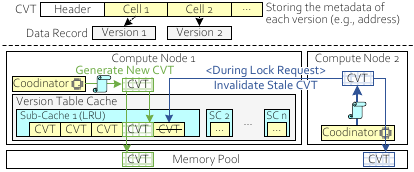}
    \captionsetup{font=small}
    \caption{The version table cache in \textsc{Lotus}.}
    \label{FIG-VTCACHE}
\end{figure}

On top of the transaction routing strategy, \textsc{Lotus} further maintains a version table cache in each CN to accelerate data retrieval without extra consistency overhead.

\noindent \textbf{Consecutive Version Table (CVT).}
\textsc{Lotus} adopts multi-version concurrency control (MVCC) for higher concurrency. In MVCC, each data record maintains multiple versions, and a version structure is used to record metadata for these versions, such as their timestamps (\textit{i.e.}, versions) and addresses. Previous work on DM~\cite{motor} uses a consecutive version table (CVT) as the version structure. As shown in Figure~\ref{FIG-VTCACHE}, the CVT stores metadata for multiple versions in a contiguous array, enabling the entire structure to be read in a single \texttt{RDMA} \texttt{READ} operation.
Similarly, \textsc{Lotus} adopts CVT as the version structure. However, this incurs performance overhead because it requires an additional \texttt{READ} to fetch the CVT and search for the correct version before reading the actual data record.

To mitigate this issue, \textsc{Lotus} maintains a version table cache in each CN, caching CVTs of data records within its managed lock range to accelerate data retrieval. This cache accelerates transaction execution by allowing the coordinators to locate the desired version and its address locally, eliminating the need for additional \texttt{RDMA} \texttt{READ}s.

\noindent \textbf{Zero Consistency Overhead.}
Combined with the hybrid transaction routing strategy (\S\ref{section_design_balancing}), maintaining consistency in the version table cache does not incur any additional overhead, as shown in Figure~\ref{FIG-VTCACHE}. This is because every write operation must first acquire a write lock, which routes its lock request to the corresponding CN. Since CVTs are exclusively modified by write operations, local transactions can synchronously update both the CVT in the memory pool and the corresponding CVT in the version table cache, thereby maintaining consistency.
For write operations initiated by remote transactions, the local coordinator invalidates the cached CVT for the relevant data when processing the lock request (line~\ref{line_invalid_cache} in Algorithm~\ref{alg_lock}). This step is crucial because the remote coordinator will directly update the CVT in the memory pool, making the cached CVT stale. By automatically invalidating cached CVTs during lock request processing, \textsc{Lotus} ensures cache consistency without introducing any additional overhead.

\noindent \textbf{LRU-Based Sub-Cache.}
\textsc{Lotus} employs a simple LRU policy for cache management. To minimize thread contention, the version table cache is further hash-partitioned into multiple independent LRU-based sub-caches, as illustrated in Figure~\ref{FIG-VTCACHE}, enabling concurrent access across different sub-caches. 

\section{Lock-First Transaction Protocol}\label{section_protocol}

\textsc{Lotus} adopts a \textit{lock-first transaction protocol} to ensure consistent and efficient transaction processing under lock disaggregation. 
This RDMA-based protocol supports MVCC and works in a simple two-phase framework: execution and commit. Notably, during a transaction, execution may occur multiple times, dynamically adding new data to the read/write sets, whereas commit happens only once at the end.
A distinctive feature of our protocol is its separation of the locking phase as the first step in each execution phase. This design effectively capitalizes on lock disaggregation: most lock operations execute locally with negligible latency, enabling early detection and abortion of conflicting transactions, thus eliminating unnecessary network overhead.

Similar to prior works~\cite{motor, ford}, our design employs a scalable timestamp service~\cite{spanner, timestampxdw, graham, sundial, hybrid-clock} deployed in the compute pool to serve all coordinators. 
Each CN maintains an address cache (\textit{i.e.}, version table address cache) to accelerate transaction execution. This address cache serves a distinct purpose compared to the previous version table cache (\S\ref{section_design_vtcache}): it stores only the addresses of CVTs and requires no active consistency maintenance, since CNs can detect stale cached addresses by validating the retrieved CVTs.

In the following section, we detail the processing phases of the lock-first transaction protocol (\S\ref{section_process_phase}) and explain how it supports different isolation levels (\S\ref{section_isolation_level}).

\subsection{Processing Phases}\label{section_process_phase}

\begin{figure}[tp]
    \centering
    \includegraphics[width=\linewidth]{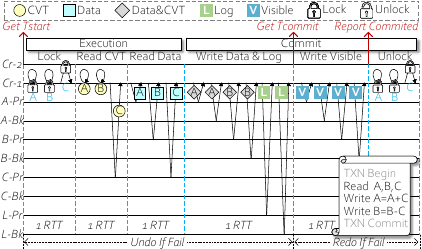}
    \captionsetup{font=small}
    \caption{The lock-first transaction protocol of \textsc{Lotus}, with a coordinator (\texttt{Cr-1}) writing data to the primary (\texttt{Pr}) and backup (\texttt{Bk}).}
    \label{FIG-PROTOCOL}
\end{figure}

Figure~\ref{FIG-PROTOCOL} illustrates the procedure of \textsc{Lotus} for processing a read-write transaction with serializability guarantee and primary-backup replication. Coordinator-1 (\texttt{Cr-1}) executes the transaction, while coordinator-2 (on another CN) processes the lock request issued from coordinator-1.  At the beginning of the transaction, the coordinator obtains a start timestamp ($T_{start}$) from the timestamp service. The read-write set is \{\texttt{A}, \texttt{B}\}, and the read-only set is \{\texttt{C}\}. Requests within one round-trip time (RTT) are issued and ACKed in parallel. 

\underline{\textit{Phase 1: Execution.}}
\textbf{1) Lock Data.} The coordinator acquires write locks for read-write data and read locks for read-only data. For local locks, it executes \texttt{CAS} instructions on the local lock table (\textit{e.g.}, \texttt{A}/\texttt{B}). For remote locks, it sends lock requests via RDMA-based RPC (\textit{e.g.}, \texttt{C}). If any lock request fails, the transaction aborts and all acquired locks are released.
\textbf{2) Read CVT}. After all locks are successfully acquired, the coordinator checks the version table cache for CVTs of locally locked data. If a cached CVT is found (\textit{e.g.}, \texttt{A}/\texttt{B}), this step is skipped. For other data, the coordinator checks the version table address cache for cached addresses. If an address is found (\textit{e.g.}, \texttt{C}), it performs an \texttt{RDMA} \texttt{READ} to read the CVT from the memory pool. Otherwise, it reads the entire CVT bucket (\textit{i.e.}, a few consecutive CVTs) and searches the corresponding CVT.
\textbf{3) Read Data.} The coordinator determines the version of the data to read by checking the CVT, selecting the largest version smaller than the start timestamp ($T_{start}$). It then reads the data record from the memory pool. If a version larger than $T_{start}$ is found, the transaction is aborted to ensure serializability, as this indicates the data has been modified and committed by another transaction.

Note that since read locks are used, \textsc{Lotus} skips the validation phase for read-only data in read-write transactions to reduce latency. This is because read locks ensure that the read-only data will not be modified during the execution, making additional validation unnecessary.

\underline{\textit{Phase 2: Commit.}}
\textbf{1) Write Data \& Log.} The coordinator writes the new CVT and data record to the memory pool via \texttt{RDMA} \texttt{WRITE}, with the version set to the 64-bit maximum value \texttt{INVISIBLE}, making the new data temporarily invisible to other coordinators.
To ensure fault tolerance (\S\ref{section_cn_recovery}), the coordinator also writes a log to its exclusive, pre-allocated memory region. As \textsc{Lotus} employs MVCC, old versions already act as the undo logs, so this log only contains some metadata (\textit{e.g.}, new data addresses), keeping the size small.
\textbf{2) Get Timestamp.} The coordinator obtains a commit timestamp ($T_{commit}$) from the timestamp service.
\textbf{3) Write Visible.} The coordinator writes the commit timestamp to all new CVTs, making the new data visible.
\textbf{4) Unlock Data.} After writing, the coordinator releases all acquired locks. For remote locks (\textit{e.g.}, \texttt{C}), it sends unlock requests via RDMA-based RPC.

Once the commit phase completes, the coordinator returns the result to the upper-layer application, indicating the transaction's completion. Note that the coordinator returns the result immediately after issuing remote unlock requests, without waiting for their completion, as the unlock process can proceed asynchronously.

\noindent \textbf{Processing Read-Only Transactions.}
The coordinator obtains a read timestamp ($T_{start}$) from the timestamp service. It then reads the CVT and selects the largest version smaller than $T_{start}$ to fetch the data record. After reading, it verifies consistency by checking the cacheline versions (see \S\ref{section_cacheline_version}). If consistent, the transaction commits; otherwise, it aborts.

\subsection{Support Different Isolation Levels}\label{section_isolation_level}

\textsc{Lotus} supports both serializability (SR) and snapshot isolation (SI) isolation levels. Under SR, concurrent transactions appear to execute sequentially, one after another. Under SI, transactions read data from a consistent snapshot at a specific point in time, unaffected by modifications made by other ongoing transactions.

\noindent \textbf{Supporting SR.}
1) For read-write transactions, \textsc{Lotus} ensures serializability by acquiring write locks for all read-write data and read locks for all read-only data. This guarantees that all data remains unchanged between the start timestamp ($T_{start}$) and the commit timestamp ($T_{commit}$). As a result, all read-write transactions can be viewed as executing serially at their commit timestamps.
2) For read-only transactions, since read-only transactions do not modify data, they do not have a commit timestamp. In MVCC, read-only transactions observe snapshot data. Their start timestamp ($T_{start}$) can be considered "movable" to fit into a serializable execution order~\cite{mvccmove}
, consistent with Motor~\cite{motor}. 
This means read-only transactions can be logically inserted between other read-write transactions, ensuring all transactions appear to execute sequentially.

\noindent \textbf{Supporting SI.}
\textsc{Lotus} ensures snapshot isolation by disabling read locks on read-only data in read-write transactions, while still requiring write locks on read-write data to prevent write-write conflicts. 
This design eliminates read-write blocking, making SI more performant than SR, a common trade-off in many industrial systems~\cite{mysqlspec, postsqlspec, oraclespec}.

\noindent \textbf{ACID Guarantees.}
\textsc{Lotus} ensures ACID properties for transactions:
1) \textbf{A}tomicity. \textsc{Lotus} maintains multiple versions of each data record, with older versions serving as undo logs to guarantee atomicity.
2) \textbf{C}onsistency. \textsc{Lotus} uses read and write locks to ensure that the state of data in the memory pool remains consistent before and after transaction execution.
3) \textbf{I}solation. \textsc{Lotus} supports both serializability and snapshot isolation to provide flexible isolation guarantees.
4) \textbf{D}urability. By varying the method and target device used for writing data and logs, \textsc{Lotus} supports using replication~\cite{primary-backup, hermes-replica}, erasure coding~\cite{carbink, effec}, 
and specialized hardware (\textit{e.g.}, UPS-backed DRAM~\cite{nocompro, motor}) to ensure data durability. 

\section{Lock-Rebuild-Free Recovery}\label{section_cn_recovery}

As \textsc{Lotus} disaggregates locks to CNs, fail-stop failures of CNs lead to the loss of locks. Restoring the old lock table during recovery would incur significant overhead. To mitigate this, \textsc{Lotus} employs a \textit{lock-rebuild-free recovery} approach, treating locks as ephemeral and avoiding their reconstruction, thereby ensuring efficient recovery. Moreover, this approach does not require specialized hardware (\textit{e.g.}, UPS-backed DRAM required by previous work~\cite{motor}).

\noindent \textbf{Failure Model.} 
We mainly discuss the fault tolerance of CN failures in \textsc{Lotus}. We consider that: 1) MNs are fault-tolerant, which can be achieved through replication~\cite{fusee, motor, hermes-replica} or erasure coding~\cite{aceso, effec, carbink}. In \textsc{Lotus}, primary-backup replication is implemented by writing data to both primary and backup MNs during the commit phase. 2) The routing layer is fault-tolerant, which can also be achieved through replication~\cite{derecho}. Since the routing layer is mostly read-only, the choice of fault-tolerance mechanism does not affect performance and is orthogonal to our work. 3) We do not consider Byzantine failures like previous researches~\cite{motor, ford, dinomo}. 

\noindent \textbf{Compute Node Failure.}\label{section_cn_recover}
\textsc{Lotus} employs a lease-based membership service~\cite{ukharon, zookeeper} to detect node failures. When a CN failure is detected, the system immediately restarts it to initiate recovery.
The loss of data in the local cache is safe. For the loss of running transactions and the lock table, \textsc{Lotus} employs the following efficient procedure to recover. 

\textit{1) Transaction recovery.} To ensure consistency, a subset of surviving coordinators scans the failed CN's write logs (persisted in the memory pool). Only transactions whose logs are fully written and whose data records have been assigned a commit timestamp (already visible) continue their commit phase. All others, whose commit phase has not yet started, are aborted to guarantee atomicity. After transaction recovery completes, surviving CNs scan their lock states and release all locks held by the failed CN by checking the holder CN ID.

\textit{2) Lock table recovery.} The failed CN's lock table is treated as ephemeral and is not rebuilt. Instead, the restarted CN begins with an empty lock table. For consistency, surviving CNs stop all transactions whose locks are held on the failed CN: only those that have already entered the commit phase continue to completion; all others are aborted, thereby releasing all locks of the failed CN. New transactions attempting to acquire locks from the failed CN are immediately aborted.

When the failed CN comes back up, it broadcasts a message confirming that all transactions in the above steps have finished. Because transaction and lock table recovery do not depend on the CN's restart and typically complete much earlier, no extra waiting is needed. The restarted CN only serves new lock requests from an empty lock table after all previously associated transactions have finished, ensuring correctness. 

\noindent \textbf{Remarks.} \textit{1) Concurrent CN failures.} Since the recovery process is decomposed into independent tasks distributed across surviving CNs (\textit{i.e.}, log scanning and local lock cleanup), concurrent failures are handled in parallel without serialization bottlenecks. \textit{2) Wait long running transactions.} In the lock table recovery, only transactions in the commit phase are waited for, which is quick and thus does not cause significant delays.
\section{Implementation}

\subsection{Memory-Side Data Store}\label{section-mem-store}

\begin{figure}[tp]
    \centering
    \includegraphics[width=\linewidth]{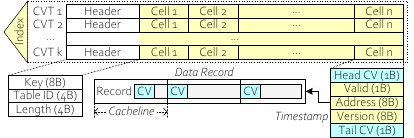}
    \captionsetup{font=small}
    \caption{\textsc{Lotus} memory store, showing one sample DB table.}
    \label{FIG-STORAGE}
\end{figure}

\noindent \textbf{RDMA-Friendly Memory Store.} In the previous MVCC-based transaction system on DM~\cite{motor}, multiple versions of a data record are organized as one full record along with delta values. This approach requires multiple small RDMA requests to reconstruct or update versions, leading to read and write amplification, wasting IOPS, and complicating protocols. To address this, \textsc{Lotus} stores each version as an independent full record and employs an efficient garbage collection mechanism to alleviate memory overhead.

As shown in Figure~\ref{FIG-STORAGE}, \textsc{Lotus} stores a consecutive version table (CVT) in the memory pool for each data record to store metadata for multiple versions. CVTs are organized by indexes. Without loss of generality, we use the hash index as a case in point, following previous studies~\cite{motor, ford}. A CVT consists of a header and $N$ cells, where each cell represents a version of the data. In the header, \texttt{Key} is the unique identifier of this record, generated by the application-specific hash function in \S\ref{section_design_sharding}. \texttt{TableID} represents the DB table this record belongs to. \texttt{Length} is the size of the data record. In each cell, \texttt{Valid} indicates whether this version is still valid. \texttt{Address} stores the 64-bit address of the corresponding record. \texttt{Version} stores the timestamp when the data is committed. During an update, the coordinator writes new data record to the memory pool and then updates the corresponding CVT cell. During a read, the coordinator reads the CVT and then reads the actual data record.

\noindent \textbf{Cacheline Version Consistency.}\label{section_cacheline_version} Since read-only transactions do not use locks, their reads may encounter inconsistencies when data is being updated. To address this, \textsc{Lotus} uses cacheline versions (CV)~\cite{farm, chime,design-guide-sync}. For CVT cells, a 1B CV is placed at their heads and tails (\textit{i.e.}, \texttt{HeadCV} and \texttt{TailCV}), while for data records, CVs are placed at the heads and cacheline boundaries. 
During updates, both the data and the corresponding cell's CVs are incremented. Readers compare the CVs in the cell and the data; mismatches indicate concurrent modifications, prompting transaction aborts.

\noindent \textbf{Lightweight Garbage Collection.} 
When updating data, the coordinator writes the new version to a free cell in the CVT. If all $N$ cells are occupied, the oldest version is overwritten to recycle its memory. In \textsc{Lotus}, timestamps are assumed to include a physical clock component, which is common in timestamp services~\cite{sundial, graham, hybrid-clock}. During writes, the coordinator also checks the timestamps of other versions. If a timestamp exceeds a threshold (\textit{e.g.}, 500ms in our implementation) relative to the local clock (with bounded drift~\cite{sundial, graham}), the cell is cleared, and its memory is reclaimed. We explore \textsc{Lotus}'s memory overhead in §~\ref{section_exp_mem_overhead} and leave more advanced GC strategy as interesting future work.

\subsection{General Optimizations}
Similar to prior works~\cite{motor, ford, dinomo, design-guide-high, sherman, understanding}, \textsc{Lotus} employs several general optimizations to maximize performance. For MNs, address alignment and huge pages are used during memory registration to accelerate memory access. For CNs, 
coroutines are enabled in coordinators to saturate the compute power of CPU cores. Small \texttt{RDMA} \texttt{WRITE}s utilize inline writes. RDMA request transmission employs doorbell batching to reduce PCIe overhead. RDMA completion queue (CQ) polling adopts selective signaling to improve performance.

\subsection{User Interface}
\textsc{Lotus} provides a user-friendly interface for applications:

\begin{table}[h]
\centering
\vspace{-0.3em}
\setlength{\abovecaptionskip}{0em}
\setlength{\belowcaptionskip}{-0.8em}
\footnotesize 
\begin{tabularx}{\linewidth}{|p{1.2cm}|X|}
\hline
\textbf{Function} & \textbf{Description} \\ \hline
\texttt{Begin()} & Start a transaction and get a start timestamp. \\ \hline
\texttt{AddRO()} & Add a data record to the read-only set. \\ \hline
\texttt{AddRW()} & Add a data record to the read-write set. \\ \hline
\texttt{Execute()} & Acquire locks, read data. \\ \hline
\texttt{Commit()} & Get a commit timestamp, write data, release locks. \\ \hline
\end{tabularx}
\label{tab:api}
\vspace{-1em} 
\end{table}

\section{Evaluation}

\subsection{Experimental Setup}\label{section_exp_setup}

\textbf{Testbed.}
We conduct all experiments on the Apt cluster of CloudLab~\cite{cloudlab} using 12 physical machines (3 MNs, 9 CNs). Each machine has two 8-core Intel E5-2650v2 CPUs, 64GB of memory, and a 56Gbps Mellanox ConnectX-3 IB RNIC, connected through a 56Gbps Mellanox SX6036G switch.

\noindent \textbf{Benchmarks.} We use a key-value store (\texttt{KVS}) as a microbenc-hmark to analyze how various factors influence \textsc{Lotus} performance. The \texttt{KVS} contains 20M KV pairs, where each key is 8B and each value is 40B. Transactions in \texttt{KVS} follow the configuration in \S\ref{section_microbenchmarks}, accessing records with varying read-write ratios. \texttt{KVS} supports both skewed and uniform access patterns, with the skewed pattern using a default Zipfian distribution ($\theta$=0.99).
We also use three widely adopted OLTP benchmarks (\texttt{TATP}~\cite{tatpspec}, \texttt{SmallBank}~\cite{smallbankspec}, and \texttt{TPCC}~\cite{tpccspec}) as macrobenchmarks to evaluate performance. \texttt{TATP} simulates telecom applications with 4 tables, where 80\% of transactions are read-only, and the maximum record size is 48B.
\texttt{SmallBank} simulates banking applications with 2 tables, where 85\% of transactions are read-write, and the record size is 16B.
\texttt{TPCC} models an ordering system with 9 tables, where 92\% of transactions are read-write, and the maximum record size is 672B. We generate 3M subscribers for \texttt{TATP}, 20M accounts for \texttt{SmallBank}, and 105 warehouses for \texttt{TPCC}.

\noindent \textbf{Comparisons.} We compare \textsc{Lotus} with Motor~\cite{motor} and FORD~\cite{ford}, two state-of-the-art transaction systems on DM. Motor supports multi-versioning, whereas FORD supports single-versioning. Although FORD is designed for PM, its one-sided RDMA based protocol is compatible with DRAM.

\noindent \textbf{Configurations.} Unless otherwise specified, we limit the lock table (\S\ref{section_design_shardlock}) to 32MB and the version table cache (\S\ref{section_design_vtcache}) to 4.5MB per CN in \textsc{Lotus}. These sizes are less than 1.3\% compared to the multi-GB working set in each MN (Figure~\ref{FIG-EXP-SYS-MEM-OVERHEAD}), making them reasonable choices. For the version table address cache, we do not impose a size limit in \textsc{Lotus}, Motor, and FORD, consistent with the previous studies~\cite{motor, ford}. For the number of versions per data in MVCC, we set it to 2 for both \textsc{Lotus} and Motor, and we explore the impact of changing this number in \S\ref{section_vtcache_size}. 
Each MN uses one CPU core only during initialization for memory registration and RDMA QP connection, and remains CPU-free during execution.
\textsc{Lotus}, Motor, and FORD all guarantee serializability (SR) and enable 3-way replication across 3 MNs (1 primary, 2 backups).

\subsection{Microbenchmarks}\label{section_microbenchmarks}

\begin{figure}[tbp]
\centering
\tiny

\subfigure[Skewed Access]{
    \includegraphics[width=0.48\linewidth]{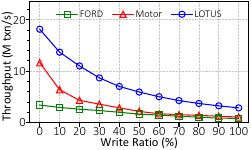}
    \label{FIG-EXP-MICRO-A}
}
\hspace{-0.8em}
\subfigure[Skewed Access]{
    \includegraphics[width=0.48\linewidth]{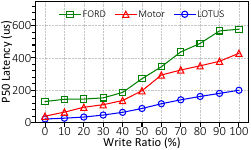}
    \label{FIG-EXP-MICRO-B}
}

\vspace{-1.4em}

\subfigure[Uniform Access]{
    \includegraphics[width=0.48\linewidth]{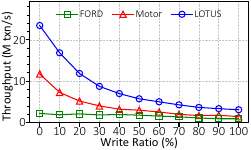}
    \label{FIG-EXP-MICRO-C}
}
\hspace{-0.8em}
\subfigure[Uniform Access]{
    \includegraphics[width=0.48\linewidth]{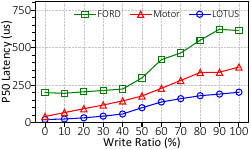}
    \label{FIG-EXP-MICRO-D}
}

\captionsetup{font=small}
\caption{The throughput and latency on \texttt{KVS} benchmark.}
\label{FIG-EXP-MICRO}
\end{figure}

The microbenchmark \texttt{KVS} includes two types of transactions: \texttt{UpdateOne} and \texttt{ReadOne}, which update and read a 40B data record, respectively. Each experiment runs for 10 seconds, during which approximately 100–200M transactions are executed. Figure~\ref{FIG-EXP-MICRO} shows the P50 latency and throughput of \textsc{Lotus}, Motor, and FORD under different read-write transaction ratios on \texttt{KVS} benchmark. Generally, as the proportion of read-write transactions increases, throughput decreases and latency rises. This is because read-write transactions involve four steps (locking, reading, writing, and unlocking), while read-only transactions require only reading. However, \textsc{Lotus} outperforms Motor and FORD across all ratios.

Figures~\ref{FIG-EXP-MICRO-A} and \ref{FIG-EXP-MICRO-B} show that under skewed accesses, \textsc{Lotus} improves throughput by 1.6–2.9$\times$ and 3.5–5.3$\times$, and reduces P50 latency by 44–64\% and 66–83\%, respectively, compared with Motor and FORD. Similarly, Figure~\ref{FIG-EXP-MICRO-C} and \ref{FIG-EXP-MICRO-D} show that under uniform accesses, \textsc{Lotus} improves throughput by 1.9–2.3$\times$ and 3.1–10.6$\times$, and reduces P50 latency by 40–66\% and 66–90\%, respectively, compared with Motor and FORD.
The performance improvement of \textsc{Lotus} in read-write transactions is due to its lock disaggregation, which bypasses the memory pool and eliminates the memory-side network bottleneck, resulting in substantial write performance gains. 
For read-only transactions, \textsc{Lotus} also demonstrates superior performance, because the version table cache accelerates some read operations. Additionally, \textsc{Lotus}'s RDMA-friendly memory store keeps a full record for each version of data, instead of a full record with deltas, reducing IOPS overhead and bandwidth amplification.

\subsection{Macrobenchmarks}\label{section_exp_macro}

\begin{figure}[tbp]
\centering
\tiny

\subfigure[\texttt{TATP}]{
    \includegraphics[width=0.48\linewidth]{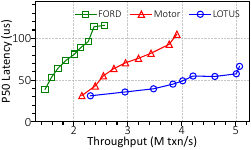}
    \label{FIG-EXP-MACRO-A}
}
\hspace{-0.8em}
\subfigure[\texttt{TATP}]{
    \includegraphics[width=0.48\linewidth]{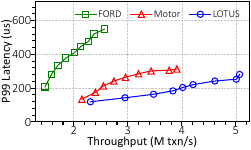}
    \label{FIG-EXP-MACRO-B}
}

\vspace{-1.4em}

\subfigure[\texttt{TPCC}]{
    \includegraphics[width=0.48\linewidth]{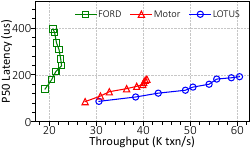}
    \label{FIG-EXP-MACRO-C}
}
\hspace{-0.8em}
\subfigure[\texttt{TPCC}]{
    \includegraphics[width=0.48\linewidth]{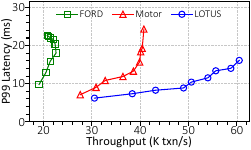}
    \label{FIG-EXP-MACRO-D}
}

\vspace{-1.4em}

\subfigure[\texttt{SmallBank}]{
    \includegraphics[width=0.48\linewidth]{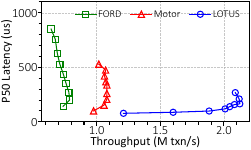}
    \label{FIG-EXP-MACRO-E}
}
\hspace{-0.8em}
\subfigure[\texttt{SmallBank}]{
    \includegraphics[width=0.48\linewidth]{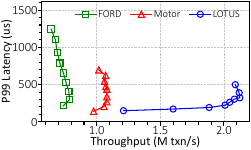}
    \label{FIG-EXP-MACRO-F}
}

\captionsetup{font=small}
\caption{The throughput and latency on \texttt{TATP}, \texttt{TPCC}, and \texttt{SmallBank} benchmarks.}
\label{FIG-EXP-MACRO}
\end{figure}

Figure~\ref{FIG-EXP-MACRO} shows the end-to-end results of \textsc{Lotus}, Motor, and FORD on macrobenchmarks: \texttt{TATP}, \texttt{TPCC}, and \texttt{SmallBank}. To generate throughput-latency curves, we increase the transaction load by running 9–90 threads and 1–6 coroutines per thread (\textit{i.e.}, 9–540 concurrent transactions).

Compared with Motor, \textsc{Lotus} improves the maximum throughput by 1.3$\times$ on \texttt{TATP}, 1.5$\times$ on \texttt{TPCC}, and 2.1$\times$ on \texttt{SmallBank}, respectively. For latency under the maximum throughput, \textsc{Lotus} reduces P50 (P99) latency by 36.7\% (10.4\%) on \texttt{TATP}, -5.2\% (33.8\%) on \texttt{TPCC}, and 49.4\% (28.5\%) on \texttt{SmallBank}. Motor performs both lock and data operations in the memory pool, leading to network contention. In contrast, \textsc{Lotus} distributes locks across CNs, resolving lock operations within the compute pool and thus eliminating this contention. 
The significant improvement on \texttt{SmallBank} compared to \texttt{TATP} and \texttt{TPCC} is due to \texttt{SmallBank}'s workload characteristics: data records are small (16B), and 85\% of transactions are read-write. This generates a large number of small-sized RDMA requests, making the performance more sensitive to IOPS-bound. \textsc{Lotus}'s lock disaggregation reduces the number of RDMA requests sent to the memory pool, especially avoiding expensive \texttt{RDMA} \texttt{CAS} operations, therefore greatly improving IOPS efficiency. On \texttt{TPCC}, \textsc{Lotus} has a slightly higher P50 latency because its write-visible step adds one RTT compared to Motor, since \textsc{Lotus} does not rely on UPS-backed DRAM for fault tolerance as Motor does.

Compared with FORD, \textsc{Lotus} improves the maximum throughput by 2.0$\times$ on \texttt{TATP}, 2.9$\times$ on \texttt{TPCC}, and 3.3$\times$ on \texttt{SmallBank}, while reducing P50 (P99) latency by 42.5\% (48.9\%) on \texttt{TATP}, 51.3\% (29.1\%) on \texttt{TPCC}, and 68.5\% (59.9\%) on \texttt{SmallBank}. FORD is limited by its single-versioning design, where data being written cannot be read simultaneously, and even read-only transactions require validation before commit, leading to lower performance.

\subsection{System-Level Analysis}\label{section-exp-sys-level}

\begin{figure}[tp]
    \centering
    \includegraphics[width=\linewidth]{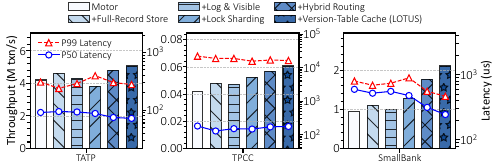}
    \captionsetup{font=small}
    \caption{The ablation study on \textsc{Lotus}.}
    \label{FIG-EXP-SYS-ABL}
\end{figure}

\noindent \textbf{Ablation Study.}
Figure~\ref{FIG-EXP-SYS-ABL} presents the ablation study of \textsc{Lotus}, showing the step-by-step impact of adding \textsc{Lotus} performance-related components, compared to Motor.
\textit{\textbf{+Full Record Store}} changes the data layout from Motor’s (one full record plus delta values) to \textsc{Lotus}’s (one full record for each version), saving bandwidth and IOPS. This improves throughput by 9.0\%, 14.2\%, and 13.9\% on \texttt{TATP}, \texttt{TPCC}, and \texttt{SmallBank}, respectively. 
\textit{\textbf{+Log \& Visible}} adds log and visible steps to the protocol, eliminating the reliance on UPS-backed DRAM for fault tolerance. This leads to a slight throughput change of -6.8\%, -1.1\%, and -8.9\% on \texttt{TATP}, \texttt{TPCC}, and \texttt{SmallBank}, respectively.
\textit{\textbf{+Lock Sharding}} distributes locks across CNs instead of MNs, allowing lock operations to bypass MNs. This improves throughput by 9.9\% on \texttt{TPCC} and 29.7\% on \texttt{SmallBank}. However, on \texttt{TATP}, throughput changes by -10.8\% because the high volume of read-only transactions saturates CN CPUs, leaving fewer cycles for handling RPCs introduced by sharding, as we choose not to consume additional CPUs for RPCs. This also increases P99 latency (\textit{e.g.}, 21.1\% on \texttt{SmallBank}) due to the scattered locks, which requires transactions to communicate with multiple CNs and thus leads to higher tail latency. 
\textit{\textbf{+Two-Level Load Balancing}} improves load balancing and directs read-write transactions to the CN where most locks reside, facilitating efficient local lock operations. 
This increases throughput by 25.7\% on \texttt{TATP}, 8.4\% on \texttt{TPCC}, and 36.5\% on \texttt{SmallBank}.
\textit{\textbf{+Version Table Cache}} further improves throughput by 5.9\%, 7.6\%, and 20.1\%, respectively, as it accelerates data retrieval.

\begin{figure}[tp]
    \centering
    \includegraphics[width=\linewidth]{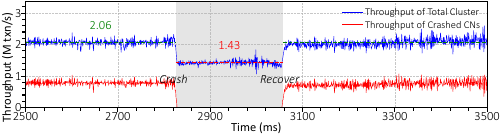}
    \captionsetup{font=small}
    \caption{The effect of CN crash on \textsc{Lotus}.}
    \label{FIG-EXP-SYS-CRASH}
\end{figure}

\noindent \textbf{Recovery of CN Crash.}\label{section_exp_cn_recover} We leverage \texttt{SmallBank} to demonstrate \textsc{Lotus}'s CN crash recovery capability, measuring throughput at 1ms intervals. During the test, we intentionally trigger the simultaneous crash of 3 CNs and immediately initiate recovery. 
Surviving CNs immediately follow the recovery process outlined in \S\ref{section_cn_recover}, without waiting for the crashed CNs to restart. The restarted CNs subsequently re-register RDMA memory regions and re-establish QP connections.
This brief disruption is illustrated in Figure~\ref{FIG-EXP-SYS-CRASH}. Around the 3-second mark, the crash causes a 30.6\% drop (\textit{i.e.}, from 2.06 to 1.43 Mtxn/s) in the total cluster throughput, but recovery completes within 233ms, which we consider acceptable~\cite{ramcloud-recover}. 

\begin{figure}[tbp]
    \centering
    \begin{minipage}[t]{0.27\linewidth}
      \centering
      \includegraphics[width=\textwidth]{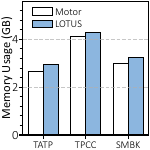}
      \captionsetup{font=small}
      \caption{Memory overhead.}
      \label{FIG-EXP-SYS-MEM-OVERHEAD}
    \end{minipage}
    \hspace{0em}
    \begin{minipage}[t]{0.71\linewidth}
      \centering
      \includegraphics[width=\textwidth]{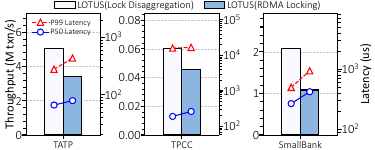}
      \captionsetup{font=small}
      \caption{Comparison with the idealized RDMA locking.}
      \label{FIG-EXP-SYS-DECLOCK}
    \end{minipage}
\end{figure}

\begin{figure}[t]
    \centering
    \begin{minipage}[t]{0.49\linewidth}
      \centering
      \includegraphics[width=\textwidth]{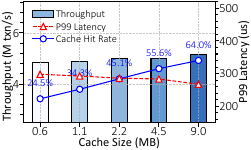}
      \captionsetup{font=small}
      \caption{Performance on \texttt{TATP} under different cache sizes.}
      \label{FIG-EXP-SENSE-CACHE-SIZE}
    \end{minipage}
    \hspace{0em}
    \begin{minipage}[t]{0.49\linewidth}
      \centering
      \includegraphics[width=\textwidth]{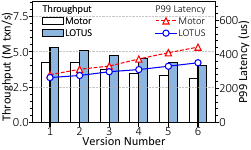}
      \captionsetup{font=small}
      \caption{Performance on \texttt{TATP} under different version numbers.}
      \label{FIG-EXP-SENSE-VCELL-NUM-A}
    \end{minipage}
\end{figure}

\noindent \textbf{Memory Overhead.}\label{section_exp_mem_overhead} To demonstrate \textsc{Lotus}'s memory overhead, we compare it with Motor, another transaction system on DM that supports MVCC, by measuring per-MN memory usage after 60 seconds of running different benchmarks: \texttt{TATP} with 3M subscribers, \texttt{TPCC} with 105 warehouses, and \texttt{SmallBank} with 20M accounts. The results are shown in Figure~\ref{FIG-EXP-SYS-MEM-OVERHEAD}. 
Even though \textsc{Lotus} stores a full copy of the data record for each version to accelerate data reads and updates, its memory overhead remains only 10.3\%, 4.7\%, and 8.5\% higher than Motor's for \texttt{TATP}, \texttt{TPCC}, and \texttt{SmallBank}, respectively. This is primarily attributed to \textsc{Lotus}'s highly efficient garbage collection mechanism.
Overall, in light of \textsc{Lotus}’s significant performance improvements, we deem the additional memory overhead acceptable.

\noindent \textbf{Benefit of Lock Disaggregation vs. RDMA Locking.} We compare \textsc{Lotus} with an idealized RDMA lock, modeled after a concurrent work~\cite{declock}. In this model, each CN maintains a local counter per lock, incrementing it on acquire and decrementing on release. An \texttt{RDMA} \texttt{FAA} is issued to the MNs only when the counter transitions between $0\rightarrow1$ or $1\rightarrow0$, simulating global ownership transfer. This model omits queues and notifications, representing a strict upper bound on performance.
As shown in Figure~\ref{FIG-EXP-SYS-DECLOCK}, \textsc{Lotus} outperforms this idealized baseline by 1.3–1.9$\times$ in throughput. While these locks~\cite{shiftlock, declock} reduce RDMA traffic via direct CN-to-CN notifications, they still incur RNIC contention on the MNs due to \texttt{RDMA} \texttt{Atomic} operations required to manage global states.

\subsection{Sensitivity Analysis}\label{section_exp_sensitivity}

\begin{figure}[t]
    \centering
    \begin{minipage}[t]{0.49\linewidth}
      \centering
      \includegraphics[width=\textwidth]{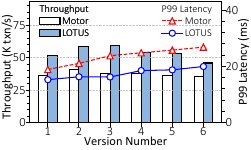}
      \captionsetup{font=small}
      \caption{Performance on \texttt{TPCC} under different version numbers.}
      \label{FIG-EXP-SENSE-VCELL-NUM-B}
    \end{minipage}
    \hspace{0em}
    \begin{minipage}[t]{0.49\linewidth}
      \centering
      \includegraphics[width=\textwidth]{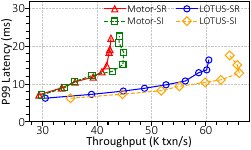}
      \captionsetup{font=small}
      \caption{Performance on \texttt{TPCC} under different isolation levels.}
      \label{FIG-EXP-SENSE-ISO}
    \end{minipage}
\end{figure}

\begin{figure}[t]
    \centering
    \begin{minipage}[t]{0.49\linewidth}
      \centering
      \includegraphics[width=\textwidth]{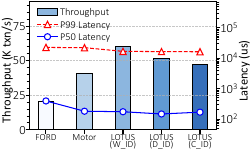}
      \captionsetup{font=small}
      \caption{Performance on \texttt{TPCC} under different critical fields.}
      \label{FIG-EXP-SENSE-CRITICAL-FIELD}
    \end{minipage}
    \hspace{0em}
    \begin{minipage}[t]{0.49\linewidth}
      \centering
      \includegraphics[width=\textwidth]{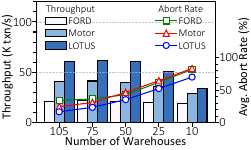}
      \captionsetup{font=small}
      \caption{Performance on \texttt{TPCC} under different warehouse counts.}
      \label{FIG-EXP-SENSE-WAREHOUSE-COUNT}
    \end{minipage}
\end{figure}

\noindent \textbf{Impact of Version Table Cache Size.}\label{section_vtcache_size} In previous tests, we fix the version table cache size in each CN to a maximum of 64K version tables (about 4.5MB). As shown in Figure~\ref{FIG-EXP-SENSE-CACHE-SIZE}, we vary its size to evaluate its impact on \texttt{TATP}. As the cache size increases, both cache hit rate and throughput improve, while P99 latency decreases. This is because the version table cache enables local CVT reads, saving an RTT and accelerating data retrieval, thereby enhancing performance.

\noindent \textbf{Impact of Version Number.} We vary the number of versions maintained per data in \textsc{Lotus} and Motor to observe its impact on \texttt{TATP} and \texttt{TPCC}, with results shown in Figures~\ref{FIG-EXP-SENSE-VCELL-NUM-A} and \ref{FIG-EXP-SENSE-VCELL-NUM-B}. \textsc{Lotus} outperforms Motor across all version numbers. For \texttt{TATP}, performance declines as the number of versions increases, as larger version table sizes lead to higher bandwidth overhead. For \texttt{TPCC}, the peak throughput is achieved at 3 versions for \textsc{Lotus} and at 2 versions for Motor, while P99 latency gradually rises with more versions. Initially, increasing versions reduces conflicts between read-write and read-only transactions, significantly lowering abort rates (\textit{e.g.}, transaction \texttt{StockLevel}'s abort rate drops from 51.3\% to 4.4\%), boosting throughput. Based on these findings, we set the default version number to 2. 

\noindent \textbf{Impact of Isolation Level.} \textsc{Lotus} and Motor both support SR and SI isolation levels, with all previous tests using SR. We evaluate their impact using \texttt{TPCC}, presented in Figure~\ref{FIG-EXP-SENSE-ISO}. \textsc{Lotus} outperforms Motor across all isolation levels. Specifically, \textsc{Lotus} in SI mode achieves 9.3\% higher maximum throughput than in SR mode, as SI eliminates read locks for read-only data, increasing concurrency. 

\noindent \textbf{Impact of Critical Field Choice.} We vary the critical field used in \textsc{Lotus} (\S\ref{section_design_sharding}) to examine its impact on \texttt{TPCC}. Specifically, \texttt{W\_ID} uses the warehouse ID as the critical field (default), \texttt{D\_ID} uses the district ID, and \texttt{C\_ID} uses the customer ID. The results are shown in Figure~\ref{FIG-EXP-SENSE-CRITICAL-FIELD}. We observe that \textsc{Lotus} does not require a perfect critical field: even a suboptimal choice still delivers performance advantages because it avoids MN‑side \texttt{RDMA} \texttt{CAS} and its associated bottleneck.

\noindent \textbf{Impact of Contention Level.} 
In the previous experiments, \texttt{TPCC} uses 105 warehouses. We reduce the warehouse count to increase contention, as shown in Figure~\ref{FIG-EXP-SENSE-WAREHOUSE-COUNT}. As contention increases, the average transaction abort rate rises. Nevertheless, \textsc{Lotus} consistently delivers the highest throughput and the lowest abort rates, demonstrating its robustness.

\section{Related Work}

\noindent \textbf{Disaggregated Memory.} Disaggregated memory (DM) has recently gained significant attention in the literature, aiming to address the imbalance between compute and memory resources. This includes advancements in indexes~\cite{sherman, smart-tree, race, rolex, chime, dex}, key-value stores~\cite{lolkv, clover, ditto, xstore, finemem, swarm}, transactions~\cite{motor, ford, hdtx, timestampxdw, hybrid-better}, networking~\cite{network-require, in-mind, mira-compiler, rdmalimiter}, operating systems~\cite{legoos, krcore, infiniswap, fastswap, clio, smartren, taleoftwo}, locks~\cite{shiftlock, citron, declock, dslr, fisslock}, replication~\cite{hermes-replica, fusee, tailwind, splitft}, erasure coding~\cite{aceso, hydra, carbink}, and LLM inferencing~\cite{splitwise, distserve, mooncacke, memserve}. \textsc{Lotus} focuses on optimizing the performance of transaction systems on DM.

\noindent \textbf{RDMA-Based Transactions.} 
RDMA-based transaction systems can be categorized into monolithic architectures~\cite{timestampxdw, hybrid-better, fast-gen-txn, farm, nocompro, FaSST, opacity, tigon} and disaggregated architectures~\cite{ford, motor, hdtx}.
Existing disaggregated solutions co-locate locks and data in the memory pool, which restricts performance. \textsc{Lotus} addresses this limitation via lock disaggregation.
% HDTX~\cite{hdtx} is a concurrent work with \textsc{Lotus}. 
HDTX~\cite{hdtx} reduces latency by offloading more processing to MN RNICs, but this aggravates the MN bottleneck and relies on the experimental RDMA feature cross-channel, which is no longer supported~\cite{crosschanelspec}.
In contrast, \textsc{Lotus} reduces the load on MN RNICs and does not depend on any experimental features.

\noindent \textbf{RDMA-Based Lock Services.} General-purpose lock services such as DSLR~\cite{dslr}, ShiftLock~\cite{shiftlock}, and DecLock~\cite{declock} provide richer functionality (\textit{e.g.}, queuing). They keep lock states in MNs and rely on \texttt{RDMA} \texttt{Atomic} for both lock acquisition and release. 
In contrast, \textsc{Lotus} adopts a specialized design for transaction processing. It treats locks as ephemeral entities managed within CNs, avoiding the overhead of \texttt{RDMA} \texttt{Atomic}.

\noindent \textbf{Ownership Partitioning.} Ownership partitioning is widely used in indexes, key-value stores, and transaction systems to improve performance, as seen in DINOMO~\cite{dinomo}, DEX~\cite{dex}, and ZEUS~\cite{zeus}. 
DINOMO, a key-value store on DM, partitions key ownership across CNs to enable efficient caching. DEX, a B+ tree index on DM, partitions subtrees across CNs for local index processing. ZEUS, a distributed transaction system based on monolithic architecture, uses dynamic partitioning to transfer data ownership to local nodes, enabling localized transaction execution. 
In contrast, \textsc{Lotus}, a distributed transaction system on DM, partitions only lock ownership, aligning well with DM characteristics. 

\section{Conclusion}

This paper presents \textsc{Lotus}, a scalable distributed transaction system on DM. To eliminate  the memory-side RNIC bottleneck, \textsc{Lotus} decouples locks from data and handles locks bypassing the memory pool. Experiments show it outperforms state-of-the-art systems in both throughput and latency.

\bibliographystyle{plain}
\bibliography{LOTUS.bib}

@inproceedings {infiniswap,
author = {Juncheng Gu and Youngmoon Lee and Yiwen Zhang and Mosharaf Chowdhury and Kang G. Shin},
title = {Efficient Memory Disaggregation with Infiniswap},
booktitle = {14th USENIX Symposium on Networked Systems Design and Implementation (NSDI 17)},
year = {2017},
isbn = {978-1-931971-37-9},
address = {Boston, MA},
pages = {649--667},
publisher = {USENIX Association},
month = mar
}

@inproceedings{bladeser,
author = {Lim, Kevin and Chang, Jichuan and Mudge, Trevor and Ranganathan, Parthasarathy and Reinhardt, Steven K. and Wenisch, Thomas F.},
title = {Disaggregated memory for expansion and sharing in blade servers},
year = {2009},
isbn = {9781605585260},
publisher = {Association for Computing Machinery},
address = {Austin, TX, USA},
booktitle = {Proceedings of the 36th Annual International Symposium on Computer Architecture},
pages = {267–278},
series = {ISCA '09}
}

@inproceedings {legoos,
author = {Yizhou Shan and Yutong Huang and Yilun Chen and Yiying Zhang},
title = {{LegoOS}: A Disseminated, Distributed {OS} for Hardware Resource Disaggregation},
booktitle = {13th USENIX Symposium on Operating Systems Design and Implementation (OSDI 18)},
year = {2018},
isbn = {978-1-939133-08-3},
address = {Carlsbad, CA},
pages = {69--87},
publisher = {USENIX Association},
month = oct
}

@inproceedings {semeru,
author = {Chenxi Wang and Haoran Ma and Shi Liu and Yuanqi Li and Zhenyuan Ruan and Khanh Nguyen and Michael D. Bond and Ravi Netravali and Miryung Kim and Guoqing Harry Xu},
title = {Semeru: A {Memory-Disaggregated} Managed Runtime},
booktitle = {14th USENIX Symposium on Operating Systems Design and Implementation (OSDI 20)},
year = {2020},
isbn = {978-1-939133-19-9},
pages = {261--280},
publisher = {USENIX Association},
month = nov
}

@inproceedings{cxl-shm,
author = {Zhang, Mingxing and Ma, Teng and Hua, Jinqi and Liu, Zheng and Chen, Kang and Ding, Ning and Du, Fan and Jiang, Jinlei and Ma, Tao and Wu, Yongwei},
title = {Partial Failure Resilient Memory Management System for (CXL-based) Distributed Shared Memory},
year = {2023},
isbn = {9798400702297},
publisher = {Association for Computing Machinery},
address = {Koblenz, Germany},
booktitle = {Proceedings of the 29th Symposium on Operating Systems Principles},
pages = {658–674},
series = {SOSP '23}
}

@misc{rdmaspec,
  author       = {Nvidia},
  title        = {RDMA Aware Networks Programming User Manual},
  year         = {2025},
  url          = {},
  note         = {\url{https://docs.nvidia.com/rdma-aware-networks-programming-user-manual-1-7.pdf}}
}

@misc{cxlspec,
  author       = {CXL},
  title        = {Compute Express Link 3.0},
  year         = {2025},
  url          = {},
  note         = {\url{https://computeexpresslink.org/wp-content/uploads/2024/02/CXL-3.0-Specification.pdf}}
}

@misc{tpccspec,
  author       = {TPC-C},
  title        = {TPC-C is an On-Line Transaction Processing Benchmark},
  year         = {2025},
  url          = {},
  note         = {\url{http://www.tpc.org/tpcc/}}
}

@misc{nvlinkspec,
  author       = {Nvidia},
  title        = {NVLink and NVLink Switch},
  year         = {2025},
  url          = {},
  note         = {\url{https://www.nvidia.com/en-us/data-center/nvlink/}}
}

@misc{crosschanelspec,
  author       = {Nvidia},
  title        = {Cross Channel},
  year         = {2025},
  url          = {},
  note         = {\url{https://docs.nvidia.com/networking/display/rdmacore50/cross+channel}}
}

@inproceedings {mooncacke,
author = {Ruoyu Qin and Zheming Li and Weiran He and Jialei Cui and Feng Ren and Mingxing Zhang and Yongwei Wu and Weimin Zheng and Xinran Xu},
title = {Mooncake: Trading More Storage for Less Computation {\textemdash} A {KVCache-centric} Architecture for Serving {LLM} Chatbot},
booktitle = {23rd USENIX Conference on File and Storage Technologies (FAST 25)},
year = {2025},
isbn = {978-1-939133-45-8},
address = {Santa Clara, CA},
pages = {155--170},
publisher = {USENIX Association},
month = feb
}

@inproceedings {ford,
author = {Ming Zhang and Yu Hua and Pengfei Zuo and Lurong Liu},
title = {{FORD}: Fast One-sided {RDMA-based} Distributed Transactions for Disaggregated Persistent Memory},
booktitle = {20th USENIX Conference on File and Storage Technologies (FAST 22)},
year = {2022},
isbn = {978-1-939133-26-7},
address = {Santa Clara, CA},
pages = {51--68},
publisher = {USENIX Association},
month = feb
}

@inproceedings {motor,
author = {Ming Zhang and Yu Hua and Zhijun Yang},
title = {Motor: Enabling {Multi-Versioning} for Distributed Transactions on Disaggregated Memory},
booktitle = {18th USENIX Symposium on Operating Systems Design and Implementation (OSDI 24)},
year = {2024},
isbn = {978-1-939133-40-3},
address = {Santa Clara, CA},
pages = {801--819},
publisher = {USENIX Association},
month = jul
}

@inproceedings {understanding,
author = {Xinhao Kong and Jingrong Chen and Wei Bai and Yechen Xu and Mahmoud Elhaddad and Shachar Raindel and Jitendra Padhye and Alvin R. Lebeck and Danyang Zhuo},
title = {Understanding {RDMA} Microarchitecture Resources for Performance Isolation},
booktitle = {20th USENIX Symposium on Networked Systems Design and Implementation (NSDI 23)},
year = {2023},
isbn = {978-1-939133-33-5},
address = {Boston, MA},
pages = {31--48},
publisher = {USENIX Association},
month = apr
}

@inproceedings{zeus,
author = {Katsarakis, Antonios and Ma, Yijun and Tan, Zhaowei and Bainbridge, Andrew and Balkwill, Matthew and Dragojevic, Aleksandar and Grot, Boris and Radunovic, Bozidar and Zhang, Yongguang},
title = {Zeus: locality-aware distributed transactions},
year = {2021},
isbn = {9781450383349},
publisher = {Association for Computing Machinery},
address = {Online Event, United Kingdom},
booktitle = {Proceedings of the Sixteenth European Conference on Computer Systems},
pages = {145–161},
numpages = {17},
series = {EuroSys '21}
}

@inproceedings{non-2pc,
author = {Lin, Qian and Chang, Pengfei and Chen, Gang and Ooi, Beng Chin and Tan, Kian-Lee and Wang, Zhengkui},
title = {Towards a Non-2PC Transaction Management in Distributed Database Systems},
year = {2016},
isbn = {9781450335317},
publisher = {Association for Computing Machinery},


booktitle = {Proceedings of the 2016 International Conference on Management of Data},
pages = {1659–1674},
numpages = {16},
address = {San Francisco, California, USA},
series = {SIGMOD '16}
}

@inproceedings {distserve,
author = {Yinmin Zhong and Shengyu Liu and Junda Chen and Jianbo Hu and Yibo Zhu and Xuanzhe Liu and Xin Jin and Hao Zhang},
title = {{DistServe}: Disaggregating Prefill and Decoding for Goodput-optimized Large Language Model Serving},
booktitle = {18th USENIX Symposium on Operating Systems Design and Implementation (OSDI 24)},
year = {2024},
isbn = {978-1-939133-40-3},
address = {Santa Clara, CA},
pages = {193--210},
publisher = {USENIX Association},
month = jul
}

@INPROCEEDINGS{splitwise,
  author={Patel, Pratyush and Choukse, Esha and Zhang, Chaojie and Shah, Aashaka and Goiri, Íñigo and Maleki, Saeed and Bianchini, Ricardo},
  booktitle={2024 ACM/IEEE 51st Annual International Symposium on Computer Architecture (ISCA)}, 
  title={Splitwise: Efficient Generative LLM Inference Using Phase Splitting}, 
  year={2024},
  volume={},
  number={},
  pages={118-132}
}

@article{memserve,
  title={MemServe: Context Caching for Disaggregated LLM Serving with Elastic Memory Pool},
  author={Cunchen Hu and Heyang Huang and Junhao Hu and Jiang Xu and Xusheng Chen and Tao Xie and Chenxi Wang and Sa Wang and Yungang Bao and Ninghui Sun and Yizhou Shan},
  year={abs/2406.17565, 2024},
  journal={ArXiv},
}

@inproceedings {cachedatten,
author = {Bin Gao and Zhuomin He and Puru Sharma and Qingxuan Kang and Djordje Jevdjic and Junbo Deng and Xingkun Yang and Zhou Yu and Pengfei Zuo},
title = {{Cost-Efficient} Large Language Model Serving for Multi-turn Conversations with {CachedAttention}},
booktitle = {2024 USENIX Annual Technical Conference (USENIX ATC 24)},
year = {2024},
isbn = {978-1-939133-41-0},
address = {Santa Clara, CA},
pages = {111--126},
publisher = {USENIX Association},
month = jul
}

@inproceedings{polarDB,
author = {Cao, Wei and Zhang, Yingqiang and Yang, Xinjun and Li, Feifei and Wang, Sheng and Hu, Qingda and Cheng, Xuntao and Chen, Zongzhi and Liu, Zhenjun and Fang, Jing and Wang, Bo and Wang, Yuhui and Sun, Haiqing and Yang, Ze and Cheng, Zhushi and Chen, Sen and Wu, Jian and Hu, Wei and Zhao, Jianwei and Gao, Yusong and Cai, Songlu and Zhang, Yunyang and Tong, Jiawang},
title = {PolarDB Serverless: A Cloud Native Database for Disaggregated Data Centers},
year = {2021},
isbn = {9781450383431},
publisher = {Association for Computing Machinery},
address = {New York, NY, USA},
booktitle = {Proceedings of the 2021 International Conference on Management of Data},
pages = {2477–2489},
numpages = {13},
location = {Virtual Event, China},
series = {SIGMOD '21}
}

@inproceedings{aurora,
author = {Verbitski, Alexandre and Gupta, Anurag and Saha, Debanjan and Brahmadesam, Murali and Gupta, Kamal and Mittal, Raman and Krishnamurthy, Sailesh and Maurice, Sandor and Kharatishvili, Tengiz and Bao, Xiaofeng},
title = {Amazon Aurora: Design Considerations for High Throughput Cloud-Native Relational Databases},
year = {2017},
isbn = {9781450341974},
publisher = {Association for Computing Machinery},
address = {Chicago, Illinois, USA},
booktitle = {Proceedings of the 2017 ACM International Conference on Management of Data},
pages = {1041–1052},
numpages = {12},
series = {SIGMOD '17}
}

@inproceedings{ditto,
author = {Shen, Jiacheng and Zuo, Pengfei and Luo, Xuchuan and Su, Yuxin and Gu, Jiazhen and Feng, Hao and Zhou, Yangfan and Lyu, Michael R.},
title = {Ditto: An Elastic and Adaptive Memory-Disaggregated Caching System},
year = {2023},
isbn = {9798400702297},
publisher = {Association for Computing Machinery},
address = {New York, NY, USA},
booktitle = {Proceedings of the 29th Symposium on Operating Systems Principles},
pages = {675–691},
numpages = {17},
location = {Koblenz, Germany},
series = {SOSP '23}
}

@inproceedings {fusee,
author = {Jiacheng Shen and Pengfei Zuo and Xuchuan Luo and Tianyi Yang and Yuxin Su and Yangfan Zhou and Michael R. Lyu},
title = {{FUSEE}: A Fully {Memory-Disaggregated} {Key-Value} Store},
booktitle = {21st USENIX Conference on File and Storage Technologies (FAST 23)},
year = {2023},
isbn = {978-1-939133-32-8},
address = {Santa Clara, CA},
pages = {81--98},
publisher = {USENIX Association},
month = feb
}

@inproceedings {smart-tree,
author = {Xuchuan Luo and Pengfei Zuo and Jiacheng Shen and Jiazhen Gu and Xin Wang and Michael R. Lyu and Yangfan Zhou},
title = {{SMART}: A {High-Performance} Adaptive Radix Tree for Disaggregated Memory},
booktitle = {17th USENIX Symposium on Operating Systems Design and Implementation (OSDI 23)},
year = {2023},
isbn = {978-1-939133-34-2},
address = {Boston, MA},
pages = {553--571},
publisher = {USENIX Association},
month = jul
}

@inproceedings{smartren,
author = {Ren, Feng and Zhang, Mingxing and Chen, Kang and Xia, Huaxia and Chen, Zuoning and Wu, Yongwei},
title = {Scaling Up Memory Disaggregated Applications with SMART},
year = {2024},
isbn = {9798400703720},
publisher = {Association for Computing Machinery},
address = {La Jolla, CA, USA},
booktitle = {Proceedings of the 29th ACM International Conference on Architectural Support for Programming Languages and Operating Systems, Volume 1},
pages = {351–367},
numpages = {17},
series = {ASPLOS '24}
}

@inproceedings{zombie,
author = {Nitu, Vlad and Teabe, Boris and Tchana, Alain and Isci, Canturk and Hagimont, Daniel},
title = {Welcome to zombieland: practical and energy-efficient memory disaggregation in a datacenter},
year = {2018},
isbn = {9781450355841},
publisher = {Association for Computing Machinery},
address = {Porto, Portugal},
booktitle = {Proceedings of the Thirteenth EuroSys Conference},
articleno = {16},
numpages = {12},
series = {EuroSys '18}
}

@inproceedings{chime,
author = {Luo, Xuchuan and Shen, Jiacheng and Zuo, Pengfei and Wang, Xin and Lyu, Michael R. and Zhou, Yangfan},
title = {CHIME: A Cache-Efficient and High-Performance Hybrid Index on Disaggregated Memory},
year = {2024},
isbn = {9798400712517},
publisher = {Association for Computing Machinery},
address = {Austin, TX, USA},
booktitle = {Proceedings of the ACM SIGOPS 30th Symposium on Operating Systems Principles},
pages = {110–126},
numpages = {17},
series = {SOSP '24}
}

@inproceedings{aceso,
author = {Hu, Zhisheng and Zuo, Pengfei and Chen, Yizou and Wang, Chao and Hu, Junliang and Yang, Ming-Chang},
title = {Aceso: Achieving Efficient Fault Tolerance in Memory-Disaggregated Key-Value Stores},
year = {2024},
isbn = {9798400712517},
publisher = {Association for Computing Machinery},
address = {Austin, TX, USA},
booktitle = {Proceedings of the ACM SIGOPS 30th Symposium on Operating Systems Principles},
pages = {127–143},
numpages = {17},
series = {SOSP '24}
}

@article{design-guide-sync,
author = {Ziegler, Tobias and Nelson-Slivon, Jacob and Leis, Viktor and Binnig, Carsten},
title = {Design Guidelines for Correct, Efficient, and Scalable Synchronization using One-Sided RDMA},
year = {2023},
issue_date = {June 2023},
publisher = {Association for Computing Machinery},
address = {New York, NY, USA},
volume = {1},
number = {2},
journal = {Proc. ACM Manag. Data},
month = jun,
articleno = {131},
numpages = {26}
}

@inproceedings{primary-backup,
author = {Lamport, Leslie and Malkhi, Dahlia and Zhou, Lidong},
title = {Vertical paxos and primary-backup replication},
year = {2009},
isbn = {9781605583969},
publisher = {Association for Computing Machinery},
address = {Calgary, AB, Canada},
booktitle = {Proceedings of the 28th ACM Symposium on Principles of Distributed Computing},
pages = {312–313},
numpages = {2},
series = {PODC '09}
}

@inproceedings {design-guide-high,
author = {Anuj Kalia and Michael Kaminsky and David G. Andersen},
title = {Design Guidelines for High Performance {RDMA} Systems},
booktitle = {2016 USENIX Annual Technical Conference (USENIX ATC 16)},
year = {2016},
isbn = {978-1-931971-30-0},
address = {Denver, CO},
pages = {437--450},
publisher = {USENIX Association},
month = jun
}

@inproceedings {FaSST,
author = {Anuj Kalia and Michael Kaminsky and David G. Andersen},
title = {{FaSST}: Fast, Scalable and Simple Distributed Transactions with {Two-Sided} ({{{{{RDMA}}}}}) Datagram {RPCs}},
booktitle = {12th USENIX Symposium on Operating Systems Design and Implementation (OSDI 16)},
year = {2016},
isbn = {978-1-931971-33-1},
address = {Savannah, GA},
pages = {185--201},
publisher = {USENIX Association},
month = nov
}

@article{dinomo,
author = {Lee, Sekwon and Ponnapalli, Soujanya and Singhal, Sharad and Aguilera, Marcos K. and Keeton, Kimberly and Chidambaram, Vijay},
title = {DINOMO: An Elastic, Scalable, High-Performance Key-Value Store for Disaggregated Persistent Memory},
year = {2022},
issue_date = {September 2022},
publisher = {VLDB Endowment},
volume = {15},
number = {13},
issn = {2150-8097},
journal = {Proc. VLDB Endow.},
month = sep,
pages = {4023–4037},
numpages = {15}
}

@article{dex,
author = {Lu, Baotong and Huang, Kaisong and Liang, Chieh-Jan Mike and Wang, Tianzheng and Lo, Eric},
title = {DEX: Scalable Range Indexing on Disaggregated Memory},
year = {2024},
issue_date = {June 2024},
publisher = {VLDB Endowment},
volume = {17},
number = {10},
issn = {2150-8097},
journal = {Proc. VLDB Endow.},
month = jun,
pages = {2603–2616},
numpages = {14}
}

@inproceedings{dynamo,
author = {DeCandia, Giuseppe and Hastorun, Deniz and Jampani, Madan and Kakulapati, Gunavardhan and Lakshman, Avinash and Pilchin, Alex and Sivasubramanian, Swaminathan and Vosshall, Peter and Vogels, Werner},
title = {Dynamo: amazon's highly available key-value store},
year = {2007},
isbn = {9781595935915},
publisher = {Association for Computing Machinery},
address = {Stevenson, Washington, USA},
booktitle = {Proceedings of Twenty-First ACM SIGOPS Symposium on Operating Systems Principles},
pages = {205–220},
numpages = {16},
series = {SOSP '07}
}

@article{spanner,
author = {Corbett, James C. and Dean, Jeffrey and Epstein, Michael and Fikes, Andrew and Frost, Christopher and Furman, J. J. and Ghemawat, Sanjay and Gubarev, Andrey and Heiser, Christopher and Hochschild, Peter and Hsieh, Wilson and Kanthak, Sebastian and Kogan, Eugene and Li, Hongyi and Lloyd, Alexander and Melnik, Sergey and Mwaura, David and Nagle, David and Quinlan, Sean and Rao, Rajesh and Rolig, Lindsay and Saito, Yasushi and Szymaniak, Michal and Taylor, Christopher and Wang, Ruth and Woodford, Dale},
title = {Spanner: Google’s Globally Distributed Database},
year = {2013},
issue_date = {August 2013},
publisher = {Association for Computing Machinery},
address = {New York, NY, USA},
volume = {31},
number = {3},
issn = {0734-2071},
journal = {ACM Trans. Comput. Syst.},
month = aug,
articleno = {8},
numpages = {22}
}

@inproceedings {timestampxdw,
author = {Xingda Wei and Rong Chen and Haibo Chen and Zhaoguo Wang and Zhenhan Gong and Binyu Zang},
title = {Unifying Timestamp with Transaction Ordering for {MVCC} with Decentralized Scalar Timestamp},
booktitle = {18th USENIX Symposium on Networked Systems Design and Implementation (NSDI 21)},
year = {2021},
isbn = {978-1-939133-21-2},
pages = {357--372},
publisher = {USENIX Association},
month = apr
}

@inproceedings{mvccmove,
author = {Neumann, Thomas and M\"{u}hlbauer, Tobias and Kemper, Alfons},
title = {Fast Serializable Multi-Version Concurrency Control for Main-Memory Database Systems},
year = {2015},
isbn = {9781450327589},
publisher = {Association for Computing Machinery},
address = {Melbourne, Victoria, Australia},
booktitle = {Proceedings of the 2015 ACM SIGMOD International Conference on Management of Data},
pages = {677–689},
numpages = {13},
series = {SIGMOD '15}
}

@misc{mysqlspec,
  author       = {MySQL},
  title        = {Transaction isolation levels},
  year         = {2025},
  url          = {},
  note         = {\url{https://dev.mysql.com/doc/refman/8.0/en/innodb-transaction-isolation-levels.html}}
}

@misc{oraclespec,
  author       = {Oracle},
  title        = {Transaction isolation levels},
  year         = {2025},
  url          = {},
  note         = {\url{https://www.oreilly.com/library/view/java-programming-with/0596000871/0596000871_orasqlj-CHP-9-SECT-2.html}}
}

@misc{postsqlspec,
  author       = {PostgreSQL},
  title        = {Transaction isolation},
  year         = {2025},
  url          = {},
  note         = {\url{https://www.postgresql.org/docs/current/transaction-iso.html}}
}

@inproceedings{hermes-replica,
author = {Katsarakis, Antonios and Gavrielatos, Vasilis and Katebzadeh, M.R. Siavash and Joshi, Arpit and Dragojevic, Aleksandar and Grot, Boris and Nagarajan, Vijay},
title = {Hermes: A Fast, Fault-Tolerant and Linearizable Replication Protocol},
year = {2020},
isbn = {9781450371025},
publisher = {Association for Computing Machinery},
address = {Lausanne, Switzerland},
booktitle = {Proceedings of the Twenty-Fifth International Conference on Architectural Support for Programming Languages and Operating Systems},
pages = {201–217},
numpages = {17},
series = {ASPLOS '20}
}

@ARTICLE{effec,
  author={Li, Qiliang and Xu, Liangliang and Li, Yongkun and Lyu, Min and Wang, Wei and Zuo, Pengfei and Xu, Yinlong},
  journal={IEEE Transactions on Parallel and Distributed Systems}, 
  title={Enabling Efficient Erasure Coding in Disaggregated Memory Systems}, 
  year={2024},
  volume={35},
  number={1},
  pages={154-168}
}

@inproceedings{nocompro,
author = {Dragojevi\'{c}, Aleksandar and Narayanan, Dushyanth and Nightingale, Edmund B. and Renzelmann, Matthew and Shamis, Alex and Badam, Anirudh and Castro, Miguel},
title = {No compromises: distributed transactions with consistency, availability, and performance},
year = {2015},
isbn = {9781450338349},
publisher = {Association for Computing Machinery},
address = {Monterey, California},
booktitle = {Proceedings of the 25th Symposium on Operating Systems Principles},
pages = {54–70},
numpages = {17},
series = {SOSP '15}
}

@article{derecho,
author = {Jha, Sagar and Behrens, Jonathan and Gkountouvas, Theo and Milano, Mae and Song, Weijia and Tremel, Edward and Renesse, Robbert Van and Zink, Sydney and Birman, Kenneth P.},
title = {Derecho: Fast State Machine Replication for Cloud Services},
year = {2019},
issue_date = {May 2018},
publisher = {Association for Computing Machinery},
address = {New York, NY, USA},
volume = {36},
number = {2},
issn = {0734-2071},
journal = {ACM Trans. Comput. Syst.},
month = apr,
articleno = {4},
numpages = {49}
}

@inproceedings {ukharon,
author = {Rachid Guerraoui and Antoine Murat and Javier Picorel and Athanasios Xygkis and Huabing Yan and Pengfei Zuo},
title = {{uKharon}: A Membership Service for Microsecond Applications},
booktitle = {2022 USENIX Annual Technical Conference (USENIX ATC 22)},
year = {2022},
isbn = {978-1-939133-29-24},
address = {Carlsbad, CA},
pages = {101--120},
publisher = {USENIX Association},
month = jul
}

@inproceedings{zookeeper,
  author       = {Patrick Hunt and
                  Mahadev Konar and
                  Flavio Paiva Junqueira and
                  Benjamin C. Reed},
  title        = {ZooKeeper: Wait-free Coordination for Internet-scale Systems},
  booktitle    = {Proceedings of the 2010 {USENIX} Annual Technical Conference, {USENIX}
                  {ATC} 2010, Boston, MA, USA, June 23-25, 2010},
  publisher    = {{USENIX} Association},
  year         = {2010}
}

@INPROCEEDINGS{hybrid-clock,
  author={Roohitavaf, Mohammad and Demirbas, Murat and Kulkarni, Sandeep},
  booktitle={2017 IEEE 36th Symposium on Reliable Distributed Systems (SRDS)}, 
  title={CausalSpartan: Causal Consistency for Distributed Data Stores Using Hybrid Logical Clocks}, 
  year={2017},
  volume={},
  number={},
  pages={184-193}
}

@inproceedings{sherman,
author = {Wang, Qing and Lu, Youyou and Shu, Jiwu},
title = {Sherman: A Write-Optimized Distributed B+Tree Index on Disaggregated Memory},
year = {2022},
isbn = {9781450392495},
publisher = {Association for Computing Machinery},
address = {Philadelphia, PA, USA},
booktitle = {Proceedings of the 2022 International Conference on Management of Data},
pages = {1033–1048},
numpages = {16},
series = {SIGMOD '22}
}

@inproceedings {cloudlab,
author = {Dmitry Duplyakin and Robert Ricci and Aleksander Maricq and Gary Wong and Jonathon Duerig and Eric Eide and Leigh Stoller and Mike Hibler and David Johnson and Kirk Webb and Aditya Akella and Kuangching Wang and Glenn Ricart and Larry Landweber and Chip Elliott and Michael Zink and Emmanuel Cecchet and Snigdhaswin Kar and Prabodh Mishra},
title = {The Design and Operation of {CloudLab}},
booktitle = {2019 USENIX Annual Technical Conference (USENIX ATC 19)},
year = {2019},
isbn = {978-1-939133-03-8},
address = {Renton, WA},
pages = {1--14},
publisher = {USENIX Association},
month = jul
}

@inproceedings {race,
author = {Pengfei Zuo and Jiazhao Sun and Liu Yang and Shuangwu Zhang and Yu Hua},
title = {One-sided {RDMA-Conscious} Extendible Hashing for Disaggregated Memory},
booktitle = {2021 USENIX Annual Technical Conference (USENIX ATC 21)},
year = {2021},
isbn = {978-1-939133-23-6},
pages = {15--29},
publisher = {USENIX Association},
month = jul
}

@inproceedings {rolex,
author = {Pengfei Li and Yu Hua and Pengfei Zuo and Zhangyu Chen and Jiajie Sheng},
title = {{ROLEX}: A Scalable {RDMA-oriented} Learned {Key-Value} Store for Disaggregated Memory Systems},
booktitle = {21st USENIX Conference on File and Storage Technologies (FAST 23)},
year = {2023},
isbn = {978-1-939133-32-8},
address = {Santa Clara, CA},
pages = {99--114},
publisher = {USENIX Association},
month = feb
}

@inproceedings {clover,
author = {Shin-Yeh Tsai and Yizhou Shan and Yiying Zhang},
title = {Disaggregating Persistent Memory and Controlling Them Remotely: An Exploration of Passive Disaggregated {Key-Value} Stores},
booktitle = {2020 USENIX Annual Technical Conference (USENIX ATC 20)},
year = {2020},
isbn = {978-1-939133-14-4},
pages = {33--48},
publisher = {USENIX Association},
month = jul
}

@inproceedings {network-require,
author = {Peter X. Gao and Akshay Narayan and Sagar Karandikar and Joao Carreira and Sangjin Han and Rachit Agarwal and Sylvia Ratnasamy and Scott Shenker},
title = {Network Requirements for Resource Disaggregation},
booktitle = {12th USENIX Symposium on Operating Systems Design and Implementation (OSDI 16)},
year = {2016},
isbn = {978-1-931971-33-1},
address = {Savannah, GA},
pages = {249--264},
publisher = {USENIX Association},
month = nov
}

@inproceedings{in-mind,
author = {Lee, Seung-seob and Yu, Yanpeng and Tang, Yupeng and Khandelwal, Anurag and Zhong, Lin and Bhattacharjee, Abhishek},
title = {MIND: In-Network Memory Management for Disaggregated Data Centers},
year = {2021},
isbn = {9781450387095},
publisher = {Association for Computing Machinery},
address = {Virtual Event, Germany},
booktitle = {Proceedings of the ACM SIGOPS 28th Symposium on Operating Systems Principles},
pages = {488–504},
numpages = {17},
series = {SOSP '21}
}

@inproceedings {xstore,
author = {Xingda Wei and Rong Chen and Haibo Chen},
title = {Fast {RDMA-based} Ordered {Key-Value} Store using Remote Learned Cache},
booktitle = {14th USENIX Symposium on Operating Systems Design and Implementation (OSDI 20)},
year = {2020},
isbn = {978-1-939133-19-9},
pages = {117--135},
publisher = {USENIX Association},
month = nov
}

@inproceedings {hybrid-better,
author = {Xingda Wei and Zhiyuan Dong and Rong Chen and Haibo Chen},
title = {Deconstructing {RDMA-enabled} Distributed Transactions: Hybrid is Better!},
booktitle = {13th USENIX Symposium on Operating Systems Design and Implementation (OSDI 18)},
year = {2018},
isbn = {978-1-939133-08-3},
address = {Carlsbad, CA},
pages = {233--251},
publisher = {USENIX Association},
month = oct
}

@inproceedings {krcore,
author = {Xingda Wei and Fangming Lu and Rong Chen and Haibo Chen},
title = {{KRCORE}: A Microsecond-scale {RDMA} Control Plane for Elastic Computing},
booktitle = {2022 USENIX Annual Technical Conference (USENIX ATC 22)},
year = {2022},
isbn = {978-1-939133-29-42},
address = {Carlsbad, CA},
pages = {121--136},
publisher = {USENIX Association},
month = jul
}

@inproceedings{fastswap,
author = {Amaro, Emmanuel and Branner-Augmon, Christopher and Luo, Zhihong and Ousterhout, Amy and Aguilera, Marcos K. and Panda, Aurojit and Ratnasamy, Sylvia and Shenker, Scott},
title = {Can far memory improve job throughput?},
year = {2020},
isbn = {9781450368827},
publisher = {Association for Computing Machinery},
address = {Heraklion, Greece},
booktitle = {Proceedings of the Fifteenth European Conference on Computer Systems},
articleno = {14},
numpages = {16},
series = {EuroSys '20}
}

@inproceedings{mira-compiler,
author = {Guo, Zhiyuan and He, Zijian and Zhang, Yiying},
title = {Mira: A Program-Behavior-Guided Far Memory System},
year = {2023},
isbn = {9798400702297},
publisher = {Association for Computing Machinery},
address = {Koblenz, Germany},
booktitle = {Proceedings of the 29th Symposium on Operating Systems Principles},
pages = {692–708},
numpages = {17},
series = {SOSP '23}
}

@inproceedings {shiftlock,
author = {Jian Gao and Qing Wang and Jiwu Shu},
title = {{ShiftLock}: Mitigate One-sided {RDMA} Lock Contention via Handover},
booktitle = {23rd USENIX Conference on File and Storage Technologies (FAST 25)},
year = {2025},
isbn = {978-1-939133-45-8},
address = {Santa Clara, CA},
pages = {355--372},
publisher = {USENIX Association},
month = feb
}

@inproceedings {citron,
author = {Jian Gao and Youyou Lu and Minhui Xie and Qing Wang and Jiwu Shu},
title = {Citron: Distributed Range Lock Management with One-sided {RDMA}},
booktitle = {21st USENIX Conference on File and Storage Technologies (FAST 23)},
year = {2023},
isbn = {978-1-939133-32-8},
address = {Santa Clara, CA},
pages = {297--314},
publisher = {USENIX Association},
month = feb
}

@inproceedings {tailwind,
author = {Yacine Taleb and Ryan Stutsman and Gabriel Antoniu and Toni Cortes},
title = {Tailwind: Fast and Atomic {RDMA-based} Replication},
booktitle = {2018 USENIX Annual Technical Conference (USENIX ATC 18)},
year = {2018},
isbn = {978-1-939133-01-4},
address = {Boston, MA},
pages = {851--863},
publisher = {USENIX Association},
month = jul
}

@inproceedings {hydra,
author = {Youngmoon Lee and Hasan Al Maruf and Mosharaf Chowdhury and Asaf Cidon and Kang G. Shin},
title = {Hydra : Resilient and Highly Available Remote Memory},
booktitle = {20th USENIX Conference on File and Storage Technologies (FAST 22)},
year = {2022},
isbn = {978-1-939133-26-7},
address = {Santa Clara, CA},
pages = {181--198},
publisher = {USENIX Association},
month = feb
}

@inproceedings {carbink,
author = {Yang Zhou and Hassan M. G. Wassel and Sihang Liu and Jiaqi Gao and James Mickens and Minlan Yu and Chris Kennelly and Paul Turner and David E. Culler and Henry M. Levy and Amin Vahdat},
title = {Carbink: {Fault-Tolerant} Far Memory},
booktitle = {16th USENIX Symposium on Operating Systems Design and Implementation (OSDI 22)},
year = {2022},
isbn = {978-1-939133-28-1},
address = {Carlsbad, CA},
pages = {55--71},
publisher = {USENIX Association},
month = jul
}

@inproceedings{fast-gen-txn,
author = {Chen, Yanzhe and Wei, Xingda and Shi, Jiaxin and Chen, Rong and Chen, Haibo},
title = {Fast and general distributed transactions using RDMA and HTM},
year = {2016},
isbn = {9781450342407},
publisher = {Association for Computing Machinery},
address = {London, United Kingdom},
booktitle = {Proceedings of the Eleventh European Conference on Computer Systems},
articleno = {26},
numpages = {17},
series = {EuroSys '16}
}

@inproceedings {farm,
author = {Aleksandar Dragojevi{\'c} and Dushyanth Narayanan and Miguel Castro and Orion Hodson},
title = {{FaRM}: Fast Remote Memory},
booktitle = {11th USENIX Symposium on Networked Systems Design and Implementation (NSDI 14)},
year = {2014},
isbn = {978-1-931971-09-6},
address = {Seattle, WA},
pages = {401--414},
publisher = {USENIX Association},
month = apr
}

@inproceedings{opacity,
author = {Shamis, Alex and Renzelmann, Matthew and Novakovic, Stanko and Chatzopoulos, Georgios and Dragojevi\'{c}, Aleksandar and Narayanan, Dushyanth and Castro, Miguel},
title = {Fast General Distributed Transactions with Opacity},
year = {2019},
isbn = {9781450356435},
publisher = {Association for Computing Machinery},
address = {Amsterdam, Netherlands},
booktitle = {Proceedings of the 2019 International Conference on Management of Data},
pages = {433–448},
numpages = {16},
series = {SIGMOD '19}
}

@misc{perftestspec,
  author       = {Infiniband},
  title        = {OFED Perftest},
  year         = {2025},
  url          = {},
  note         = {\url{https://github.com/linux-rdma/perftest}}
}

@article{tail-scale,
author = {Dean, Jeffrey and Barroso, Luiz Andr\'{e}},
title = {The tail at scale},
year = {2013},
issue_date = {February 2013},
publisher = {Association for Computing Machinery},
address = {New York, NY, USA},
volume = {56},
number = {2},
issn = {0001-0782},
journal = {Commun. ACM},
month = feb,
pages = {74–80},
numpages = {7}
}

@inproceedings{ramcloud-recover,
author = {Ongaro, Diego and Rumble, Stephen M. and Stutsman, Ryan and Ousterhout, John and Rosenblum, Mendel},
title = {Fast crash recovery in RAMCloud},
year = {2011},
isbn = {9781450309776},
publisher = {Association for Computing Machinery},
address = {Cascais, Portugal},
booktitle = {Proceedings of the Twenty-Third ACM Symposium on Operating Systems Principles},
pages = {29–41},
numpages = {13},
series = {SOSP '11}
}

@misc{tatpspec,
  author       = {TATP},
  title        = {Telecom Application Transaction Processing Benchmark},
  year         = {2025},
  url          = {},
  note         = {\url{https://tatpbenchmark.sourceforge.net/}}
}

@misc{smallbankspec,
  author       = {SmallBank},
  title        = {SmallBank Benchmark},
  year         = {2025},
  url          = {},
  note         = {\url{https://hstore.cs.brown.edu/documentation/deployment/benchmarks/smallbank/}}
}

@inproceedings{venmo,
author = {Unger, Clive and Murthy, Dhiraj and Acker, Amelia and Arora, Ishank and Chang, Andy},
title = {Examining the evolution of mobile social payments in Venmo},
year = {2020},
isbn = {9781450376884},
publisher = {Association for Computing Machinery},
address = {Toronto, ON, Canada},
booktitle = {International Conference on Social Media and Society},
pages = {101–110},
numpages = {10},
series = {SMSociety'20}
}

@inproceedings{scale-store,
author = {Ziegler, Tobias and Binnig, Carsten and Leis, Viktor},
title = {ScaleStore: A Fast and Cost-Efficient Storage Engine using DRAM, NVMe, and RDMA},
year = {2022},
isbn = {9781450392495},
publisher = {Association for Computing Machinery},
address = {Philadelphia, PA, USA},
booktitle = {Proceedings of the 2022 International Conference on Management of Data},
pages = {685–699},
numpages = {15},
series = {SIGMOD '22}
}

@inproceedings {twitter-workload,
author = {Juncheng Yang and Yao Yue and K. V. Rashmi},
title = {A large scale analysis of hundreds of in-memory cache clusters at Twitter},
booktitle = {14th USENIX Symposium on Operating Systems Design and Implementation (OSDI 20)},
year = {2020},
isbn = {978-1-939133-19-9},
pages = {191--208},
publisher = {USENIX Association},
month = nov
}

@inproceedings {pegasus,
author = {Jialin Li and Jacob Nelson and Ellis Michael and Xin Jin and Dan R. K. Ports},
title = {Pegasus: Tolerating Skewed Workloads in Distributed Storage with {In-Network} Coherence Directories},
booktitle = {14th USENIX Symposium on Operating Systems Design and Implementation (OSDI 20)},
year = {2020},
isbn = {978-1-939133-19-9},
pages = {387--406},
publisher = {USENIX Association},
month = nov
}

@inproceedings {facebook-workload,
author = {Zhichao Cao and Siying Dong and Sagar Vemuri and David H.C. Du},
title = {Characterizing, Modeling, and Benchmarking {RocksDB} {Key-Value} Workloads at Facebook},
booktitle = {18th USENIX Conference on File and Storage Technologies (FAST 20)},
year = {2020},
isbn = {978-1-939133-12-0},
address = {Santa Clara, CA},
pages = {209--223},
url = {https://www.usenix.org/conference/fast20/presentation/cao-zhichao},
publisher = {USENIX Association},
month = feb
}

@inproceedings {fisslock,
author = {Hanze Zhang and Ke Cheng and Rong Chen and Haibo Chen},
title = {Fast and Scalable In-network Lock Management Using Lock Fission},
booktitle = {18th USENIX Symposium on Operating Systems Design and Implementation (OSDI 24)},
year = {2024},
isbn = {978-1-939133-40-3},
address = {Santa Clara, CA},
pages = {251--268},
url = {https://www.usenix.org/conference/osdi24/presentation/zhang-hanze},
publisher = {USENIX Association},
month = jul
}

@inproceedings {hdtx,
author = {Haodi Lu and Haikun Liu and Yujian Zhang and Zhuohui Duan and Xiaofei Liao and Hai Jin and Yu Zhang},
title = {Fast Distributed Transactions for RDMA-based Disaggregated Memory},
booktitle = {2025 USENIX Annual Technical Conference (USENIX ATC 25)},
year = {2025},
address = {Santa Clara, CA},
pages = {943--958},
publisher = {USENIX Association},
month = jul
}

@inproceedings {finemem,
author = {Xiaoyang Wang and Yongkun Li and Kan Wu and Wenzhe Zhu and Yuqi Li and Yinlong Xu},
title = {FineMem: Breaking the Allocation Overhead vs. Memory Waste Dilemma in Fine-Grained Disaggregated Memory Management},
booktitle = {19th USENIX Symposium on Operating Systems Design and Implementation (OSDI 25)},
year = {2025},
address = {Santa Clara, CA},
pages = {57--74},
publisher = {USENIX Association},
month = jul
}

@inproceedings{swarm,
author = {Murat, Antoine and Burgelin, Cl\'{e}ment and Xygkis, Athanasios and Zablotchi, Igor and Aguilera, Marcos Kawazoe and Guerraoui, Rachid},
title = {SWARM: Replicating Shared Disaggregated-Memory Data in No Time},
year = {2024},
isbn = {9798400712517},
publisher = {Association for Computing Machinery},
address = {Austin, TX, USA},
booktitle = {Proceedings of the ACM SIGOPS 30th Symposium on Operating Systems Principles},
pages = {24–45},
numpages = {22},
series = {SOSP '24}
}

@inproceedings{cxl-cloud-native,
author = {Yang, Xinjun and Zhang, Yingqiang and Chen, Hao and Li, Feifei and Fan, Gerry and Kong, Yang and Wang, Bo and Fang, Jing and Wang, Yuhui and Huang, Tao and Hu, Wenpu and Kao, Jim and Jiang, Jianping},
title = {Unlocking the Potential of CXL for Disaggregated Memory in Cloud-Native Databases},
year = {2025},
isbn = {9798400715648},
publisher = {Association for Computing Machinery},
address = {Berlin, Germany},
booktitle = {Companion of the 2025 International Conference on Management of Data},
pages = {689–702},
numpages = {14},
series = {SIGMOD/PODS '25}
}

@inproceedings {tigon,
author = {Yibo Huang and Haowei Chen and Newton Ni and Yan Sun and Vijay Chidambaram and Dixin Tang and Emmett Witchel},
title = {Tigon: A Distributed Database for a CXL Pod},
booktitle = {19th USENIX Symposium on Operating Systems Design and Implementation (OSDI 25)},
year = {2025},
address = {Santa Clara, CA},
pages = {57--74},
publisher = {USENIX Association},
month = jul
}

@article{declock,
  title={DecLock: A Case of Decoupled Locking for Disaggregated Memory},
  author={Hanze Zhang and Ke Cheng and Rong Chen and Xingda Wei and Haibo Chen},
  year={abs/2505.17641, 2025},
  journal={ArXiv},
}

@inproceedings{dslr,
author = {Yoon, Dong Young and Chowdhury, Mosharaf and Mozafari, Barzan},
title = {Distributed Lock Management with RDMA: Decentralization without Starvation},
year = {2018},
isbn = {9781450347037},
publisher = {Association for Computing Machinery},
address = {Houston, TX, USA},
booktitle = {Proceedings of the 2018 International Conference on Management of Data},
pages = {1571–1586},
numpages = {16},
series = {SIGMOD '18}
}

@inproceedings {sundial,
author = {Yuliang Li and Gautam Kumar and Hema Hariharan and Hassan Wassel and Peter Hochschild and Dave Platt and Simon Sabato and Minlan Yu and Nandita Dukkipati and Prashant Chandra and Amin Vahdat},
title = {Sundial: Fault-tolerant Clock Synchronization for Datacenters},
booktitle = {14th USENIX Symposium on Operating Systems Design and Implementation (OSDI 20)},
year = {2020},
isbn = {978-1-939133-19-9},
pages = {1171--1186},
url = {https://www.usenix.org/conference/osdi20/presentation/li-yuliang},
publisher = {USENIX Association},
month = nov
}

@inproceedings {graham,
author = {Ali Najafi and Michael Wei},
title = {Graham: Synchronizing Clocks by Leveraging Local Clock Properties},
booktitle = {19th USENIX Symposium on Networked Systems Design and Implementation (NSDI 22)},
year = {2022},
isbn = {978-1-939133-27-4},
address = {Renton, WA},
pages = {453--466},
url = {https://www.usenix.org/conference/nsdi22/presentation/najafi},
publisher = {USENIX Association},
month = apr
}

@inproceedings {lolkv,
author = {Ahmed Alquraan and Sreeharsha Udayashankar and Virendra Marathe and Bernard Wong and Samer Al-Kiswany},
title = {{LoLKV}: The Logless, Linearizable, {RDMA-based} {Key-Value} Storage System},
booktitle = {21st USENIX Symposium on Networked Systems Design and Implementation (NSDI 24)},
year = {2024},
isbn = {978-1-939133-39-7},
address = {Santa Clara, CA},
pages = {41-54},
url = {https://www.usenix.org/conference/nsdi24/presentation/alquraan},
publisher = {USENIX Association},
month = apr
}

@inproceedings{splitft,
author = {Luo, Xuhao and Alagappan, Ramnatthan and Ganesan, Aishwarya},
title = {SplitFT: Fault Tolerance for Disaggregated Datacenters via Remote Memory Logging},
year = {2024},
publisher = {Association for Computing Machinery},
address = {Athens, Greece},
booktitle = {Proceedings of the Nineteenth European Conference on Computer Systems},
pages = {590–607},
numpages = {18},
series = {EuroSys '24}
}

@inproceedings{cxlpond,
author = {Li, Huaicheng and Berger, Daniel S. and Hsu, Lisa and Ernst, Daniel and Zardoshti, Pantea and Novakovic, Stanko and Shah, Monish and Rajadnya, Samir and Lee, Scott and Agarwal, Ishwar and Hill, Mark D. and Fontoura, Marcus and Bianchini, Ricardo},
title = {Pond: CXL-Based Memory Pooling Systems for Cloud Platforms},
year = {2023},
publisher = {Association for Computing Machinery},
address = {Vancouver, BC, Canada},
booktitle = {Proceedings of the 28th ACM International Conference on Architectural Support for Programming Languages and Operating Systems, Volume 2},
pages = {574–587},
numpages = {14},
series = {ASPLOS 2023}
}

@inproceedings{clio,
author = {Guo, Zhiyuan and Shan, Yizhou and Luo, Xuhao and Huang, Yutong and Zhang, Yiying},
title = {Clio: a hardware-software co-designed disaggregated memory system},
year = {2022},
isbn = {9781450392051},
publisher = {Association for Computing Machinery},
address = {Lausanne, Switzerland},
booktitle = {Proceedings of the 27th ACM International Conference on Architectural Support for Programming Languages and Operating Systems},
pages = {417–433},
numpages = {17},
series = {ASPLOS '22}
}

@inproceedings{rethink,
author = {Calciu, Irina and Imran, M. Talha and Puddu, Ivan and Kashyap, Sanidhya and Maruf, Hasan Al and Mutlu, Onur and Kolli, Aasheesh},
title = {Rethinking software runtimes for disaggregated memory},
year = {2021},
publisher = {Association for Computing Machinery},
address = {Virtual, USA},
booktitle = {Proceedings of the 26th ACM International Conference on Architectural Support for Programming Languages and Operating Systems},
pages = {79–92},
numpages = {14},
series = {ASPLOS '21}
}

@inproceedings {taleoftwo,
author = {Lei Chen and Shi Liu and Chenxi Wang and Haoran Ma and Yifan Qiao and Zhe Wang and Chenggang Wu and Youyou Lu and Xiaobing Feng and Huimin Cui and Shan Lu and Harry Xu},
title = {A Tale of Two Paths: Toward a Hybrid Data Plane for Efficient {Far-Memory} Applications},
booktitle = {18th USENIX Symposium on Operating Systems Design and Implementation (OSDI 24)},
year = {2024},
isbn = {978-1-939133-40-3},
address = {Santa Clara, CA},
pages = {77--95},
publisher = {USENIX Association},
month = jul
}

@article{gaussdb,
author = {Li, Guoliang and Tian, Wengang and Zhang, Jinyu and Grosman, Ronen and Liu, Zongchao and Li, Sihao},
title = {GaussDB: A Cloud-Native Multi-Primary Database with Compute-Memory-Storage Disaggregation},
year = {2024},
issue_date = {August 2024},
publisher = {VLDB Endowment},
volume = {17},
number = {12},
issn = {2150-8097},
journal = {Proc. VLDB Endow.},
month = aug,
pages = {3786–3798},
numpages = {13}
}

@inproceedings{rdmalimiter,
author = {Wang, Zilong and Wan, Xinchen and Li, Luyang and Sun, Yijun and Xie, Peng and Wei, Xin and Ning, Qingsong and Zhang, Junxue and Chen, Kai},
title = {Fast, Scalable, and Accurate Rate Limiter for RDMA NICs},
year = {2024},
isbn = {9798400706141},
publisher = {Association for Computing Machinery},
address = {New York, NY, USA},
booktitle = {Proceedings of the ACM SIGCOMM 2024 Conference},
pages = {568–580},
numpages = {13},
location = {Sydney, NSW, Australia},
series = {ACM SIGCOMM '24}
}

@misc{mongodbspec,
  author       = {MongoDB},
  title        = {Choose a shard key},
  year         = {2025},
  url          = {},
  note         = {\url{https://www.mongodb.com/docs/manual/core/sharding-choose-a-shard-key/}}
}

@misc{dynamodbspec,
  author       = {Amazon DynamoDB},
  title        = {Best practices for designing and using partition keys effectively in DynamoDB},
  year         = {2025},
  url          = {},
  note         = {\url{https://docs.aws.amazon.com/amazondynamodb/latest/developerguide/bp-partition-key-design.html}}
}

@misc{cosmosdbspec,
  author       = {Azure Cosmos DB},
  title        = {Partitioning and horizontal scaling in Azure Cosmos DB},
  year         = {2025},
  url          = {},
  note         = {\url{https://learn.microsoft.com/en-us/azure/cosmos-db/partitioning-overview}}
}

@misc{cassandraspec,
  author       = {Apache Cassandra},
  title        = {Cassandra basics},
  year         = {2025},
  url          = {},
  note         = {\url{https://cassandra.apache.org/_/cassandra-basics.html}}
}

@inproceedings {rdmaturing,
author = {Waleed Reda and Marco Canini and Dejan Kosti{\'c} and Simon Peter},
title = {{RDMA} is Turing complete, we just did not know it yet!},
booktitle = {19th USENIX Symposium on Networked Systems Design and Implementation (NSDI 22)},
year = {2022},
isbn = {978-1-939133-27-4},
address = {Renton, WA},
pages = {71--85},
publisher = {USENIX Association},
month = apr
}

@inproceedings{oasis,
author = {Zhong, Yuhong and Berger, Daniel S. and Zardoshti, Pantea and Saurez, Enrique and Nelson, Jacob and Ports, Dan R. K. and Psistakis, Antonis and Fried, Joshua and Cidon, Asaf},
title = {Oasis: Pooling PCIe Devices Over CXL to Boost Utilization},
year = {2025},
isbn = {9798400718700},
publisher = {Association for Computing Machinery},
address = {Lotte Hotel World, Seoul, Republic of Korea},
url = {https://doi.org/10.1145/3731569.3764812},
doi = {10.1145/3731569.3764812},
booktitle = {Proceedings of the ACM SIGOPS 31st Symposium on Operating Systems Principles},
pages = {101–119},
numpages = {19},
series = {SOSP '25}
}

@inproceedings{tigatxn,
author = {Geng, Jinkun and Mu, Shuai and Sivaraman, Anirudh and Prabhakar, Balaji},
title = {Tiga: Accelerating Geo-Distributed Transactions with Synchronized Clocks},
year = {2025},
isbn = {9798400718700},
publisher = {Association for Computing Machinery},
address = {Lotte Hotel World, Seoul, Republic of Korea},
url = {https://doi.org/10.1145/3731569.3764854},
doi = {10.1145/3731569.3764854},
booktitle = {Proceedings of the ACM SIGOPS 31st Symposium on Operating Systems Principles},
pages = {555–571},
numpages = {17},
series = {SOSP '25}
}

@article{yahootrace,
author = {Cooper, Brian F. and Ramakrishnan, Raghu and Srivastava, Utkarsh and Silberstein, Adam and Bohannon, Philip and Jacobsen, Hans-Arno and Puz, Nick and Weaver, Daniel and Yerneni, Ramana},
title = {PNUTS: Yahoo!'s hosted data serving platform},
year = {2008},
issue_date = {August 2008},
publisher = {VLDB Endowment},
volume = {1},
number = {2},
issn = {2150-8097},
url = {https://doi.org/10.14778/1454159.1454167},
doi = {10.14778/1454159.1454167},
journal = {Proc. VLDB Endow.},
month = aug,
pages = {1277–1288},
numpages = {12}
}

\end{document}